\documentclass[12pt]{iopart}
\usepackage{iopams}
\usepackage{comment}
\usepackage{enumerate}
\usepackage{color}
\usepackage{graphicx}
\usepackage{subfigure}
\usepackage{hyperref}

\eqnobysec

\newcommand{\conn}[3]{\ensuremath{\Gamma^{#1}_{\ #2#3}}} 
\newcommand{\krod}[2]{\ensuremath{\delta^{#1}_{\ #2}}} 
\newcommand{\riem}[4]{\ensuremath{R^{#1}_{\ #2#3#4}}} 

\begin{document}

\title[Differential geometric analysis of radiation-particle interaction]{Differential geometric analysis of radiation-particle interaction}

\author{Kiam Heong Kwa}

\address{Department of Mathematics, The Ohio State University, Columbus, OH 43210, USA}
\eads{\mailto{khkwa@math.ohio-state.edu}, \mailto{khkwa@hotmail.com}}
\begin{abstract}
On the basis of the Lorentz equations of motion, the orbit of a charge driven by a generic E.M. field with planar symmetry is formulated and analyzed within the framework of a Lorentzian geometry with a curvature whose order of magnitude is parametrized by the radiation intensity and frequency. This reformulation leads to (i) a demonstration of the integrability of the particle motion in a plane-wave field as a result of the vanishing of the curvature, (ii) a manifestation of the parametric dependence of the dynamical response of the particle orbit to the E.M. field on the impulse factor in terms of local stability and the occurrence of parametric resonance and (iii) a mathematically precise meaning to the ponderomotive oscillation center of the charge executing oscillatory motion in a sufficiently low impulsive E.M. field, which is subsequently used to raise (iv) a discussion of the domain of applicability of the ponderomotive approximation.
\end{abstract}



\section{Introduction}
\label{section I}

The starting point for many investigations on nonlinear radiation-matter interaction has been the orbit of a single charged particle in an E.M. plane wave studied on the basis of the Lorentz equations of motion or its equivalent Lagrangian formulation. For example, see \cite{bardsley1989,bauer1995,galkin2008,hartemann1998,sarachik1970,startsev1997}. The primary aims of this paper are (1) to give an exact differential geometric formulation of the Lorentz equations of motion in an E.M. field with planar symmetry and (2) to analyze the dynamics of the orbits of charged particles in response to such an E.M. radiation within the geometrical framework. It is worth noting that the class of E.M. fields with planar symmetry includes arbitrary plane-wave fields, standing-wave fields, and E.M. fields of two-counter-propagating waves as particular instances. See (\ref{III.1}), (\ref{IV.3}), and (\ref{V.0.1}) below. It is shown that \emph{the complexity of particle dynamics can be encapsulated economically into a curvature function and associated geometrical quantities.} To demonstrate this conceptual novelty, we show that \emph{the integrability of the particle motion in an arbitrary E.M. with planar symmetry is a consequence of the vanishing of the curvature;} the widely known \emph{admission of analytic solutions by the Lorentz equations for arbitrary plane-wave fields is therefore a manifestation of the flatness of a manifold.}

Another demonstration of the conceptual novelty, which is more important, is the availability of the Jacobi equations for geodesic flows and the associated curvture features. This way \emph{the stability question of particle orbits is translated into the measure of the strength of the geodesic spread.} To illustrate this, we consider the stability of particle orbits in a standing wave field by analyzing the associated Jacobi fields and indicate the interrelation of the (in)stability of particle orbits and (non)occurene of parametric resonance. Finally, as an important application of the differential geometric framework, we employ a mathematical averaging of the Jacobi equations and the presence of parametric resonance to discuss the domain of applicability of the ponderomotive oscillation center dynamics of charged particles.

The outline of this paper is as follows. In Sec. \ref{section II}, the identification of the physical orbits of a natural mechanical system with the geodesics in the configuration space with a suitable semi-Riemannian metric is established. In Sec. \ref{section III}, the reduction of the interaction of a charged particle with a generic planar symmetic E.M. field to a natural mechanical system is made. This sets up the foundation of the differential geometric analysis of the radiation-particle interaction in this paper. 

In Sec. \ref{section IV}, the well-known integrability of the particle motion in a plane-wave field is shown as a result of the vanishing of a curvature function. In addition, the analytic solutions of the Lorentz equations for an arbitrary plane-wave field is obtained from the corresponding geodesic equations. 
 
In Sec. \ref{section V}, the stability question of the particle motion in an E.M. field of standing wave is studied. The effect of the dynamical response of a particle of charge $q$ and mass $m$ placed in an E.M. field of radiation frequency $\omega$ and field amplitude $\cal{E}$ is very often expressed by the magnitude of the dimensionless \emph{impulse factor}
\begin{equation}\label{I.1}
\eta=\frac{q{\cal E}}{m\omega},
\end{equation}
which is usually also called the electron quiver velocity if the charge is an electron or the intensity parameter \cite{gibbon2005,salamin2006}. There is no exception when the particle interacts with a generic E.M. field with planar symmetry. However, we will show in this paper that \emph{the stability of the orbit of a particle interacting with the E.M. field is not completely determined by the order of magnitude of the impulse factor $\eta$, but by the different numerical subintervals within which the value of $\eta$ lies.} In particular, a highly impulsive E.M. field ($\eta\gg1 $) may not lead to instability of orbits of particles. This, as we will show,  is a consequence of the (non)occurrence of parametric resonance in the system consisting of the particles and the E.M. field, where the impulse factor $\eta$ is the parameter of the system that determines the (non)occurence of such a parametric resonance.

We also employ the Jacobi equations to consider the applicability of the oscillation center dynamics of charged particles. The oscillation center dynamics of a charged particle in an E.M. field, governed by the so-called ponderomotive force, has been the basis of many investigations of the particle motion in response to the field \cite{bauer1995,freeman1988,gibbon2005,hartemann1998,kaplan2005,mulser2010,nicholson1983,pokrovsky2005,schmidt1979,startsev1997}. As another significant application of the differential geometric framework, we provide a mathematically precise meaning to the ponderomotive oscillation center of a charged particle executing oscillatory motion in a sufficiently low impulsive E.M. field ($\eta\ll 1$). This we do by indicating the quantitative relation of a Jacobi field and the associated ponderomotive oscillation center to the corresponding solution of a mathematically precise averaged version of the Jacobi equations. With the aid of this setup, it is shown that \emph{when a parametric resonance occurs, the ponderomotive approximation breaks down.}

Finally, in Sec. \ref{section VI}, we summarize the results of this paper.
\section{Semi-Riemannian Formulation of Natural Mechanical Systems}
\label{section II}

This is a preparatory section in which we derive the identification of the physical orbits of a natural mechanical system, a dynamical system that can be described by a Lagrangian of the form
\begin{equation}\label{II.1}
{\cal L}\left(x^A,\frac{dx^A}{d\tau}\right)=\frac{k_{BC}}{2}\frac{dx^B}{d\tau}\frac{dx^C}{d\tau}-\Phi(x^A),
\end{equation}
where $k_{AB}$ is a metric based kinetic energy tensor and $\Phi$ is a potential, all of which depend only on the coordinates $x^A$, with the geodesics in the configuration space endowed with a suitable semi-Riemannian metric. We will also recall the geodesic equations and the Jacobi equations for geodesic spread for later reference.  As a convention, all capital Latin letters A, B, C, and so on, whenever appearing as subscripts or superscripts, always run over the values $0,1,\cdots,N-1$, where $N$ is the number of degrees of freedom of the original dynamical system.

As is well-known, a natural mechanical system can also be described by the Hamiltonian
\begin{equation}\label{II.2}
{\cal H}(p_A,x^A)=\frac{k^{BC}p_Bp_C}{2}+\Phi(x^A),
\end{equation}
where $p_A=\partial{\cal L}/\partial(dx^A/d\tau)=k_{AB}dx^B/d\tau$ are the generalized momenta, and the Hamiltonian is an integral of motion. We are particularly interested in the dynamics of timelike orbits along which $k_{AB}\left(dx^A/d\tau\right)\left(dx^B/d\tau\right)=H-\Phi<0$ for a fixed Hamiltonian value ${\cal H}=H$ in the case $k_{AB}$ is Lorentzian. In this case, it can be shown that the timelike orbits are geodesics in the configuration space endowed with the metric tensor
\begin{equation}\label{II.3}
g_{AB}=-2(H-\Phi)k_{AB},
\end{equation}
called the Jacobi metric. This is a classical result for the Riemannian case, a derivation based on Hamiltonian's least action principle of which can be found in \cite{casetti1996,cerruti1996,pettini1993,pettini2007}. However, we show that it comes as a consequence of the duality of the Hamiltonian ${\cal H}$ and the orbit parameter $\tau$. Explicitly, we consider $\tau$ as an additional degree of freedom and reparametrizes the orbits by a new parameter $\lambda$ with the property that $d\tau/d\lambda>0$. Since $\int_{\tau(\lambda_0)}^{\tau(\lambda_1)}\,{\cal L}\,d\tau=\int_{\lambda_0}^{\lambda_1}\,{\cal L}\left(d\tau/d\lambda\right)\,d\lambda$, the Lagrange's equations in the $x^A$-directions for the extended Lagrangian
\begin{equation}\label{II.4}
\eqalign{
{\cal L}^e\left(x^A,\frac{dx^A}{d\lambda},\frac{d\tau}{d\lambda}\right)&={\cal L}\left(x^A,\frac{dx^A}{d\lambda}\frac{d\tau}{d\lambda}\right)\frac{d\tau}{d\lambda}\cr
&=\frac{k_{BC}}{2}\frac{dx^B}{d\lambda}\frac{dx^C}{d\lambda}\frac{d\lambda}{d\tau}-\Phi(x^A)\frac{d\tau}{d\lambda}}
\end{equation}
are equivalent to the ones for the original Lagrangian ${\cal L}$. Then since $\tau$ is cyclic with respect to ${\cal L}^e$, 
\begin{equation}\label{II.5}
\frac{\partial{\cal L}^e}{\partial(d\tau/d\lambda)}=-\frac{k_{AB}}{2}\frac{dx^A}{d\lambda}\frac{dx^B}{d\lambda}\left(\frac{d\lambda}{d\tau}\right)^2-\Phi
\end{equation}
is an integral of motion. In fact, comparing (\ref{II.2}) and (\ref{II.5}), we see that ${\cal L}^e$ has the value $-H$, so that
\begin{equation}\label{II.6}
\frac{k_{AB}}{2}\frac{dx^A}{d\lambda}\frac{dx^B}{d\lambda}\left(\frac{d\lambda}{d\tau}\right)^2+\Phi=H.
\end{equation}
Solving for 
\begin{equation}\label{II.7}
\frac{d\tau}{d\lambda}=\sqrt{\frac{k_{AB}}{2(H-\Phi)}\frac{dx^A}{d\lambda}\frac{dx^B}{d\lambda}}
\end{equation}
from (\ref{II.6}) and applying the classical Routh's procedure yields the Routhian
\begin{equation}\label{II.8}
\eqalign{{\cal L}^e_s\left(x^A,\frac{dx^A}{d\lambda}\right)&={\cal L}^e-\frac{\partial{\cal L}^e}{\partial(d\tau/d\lambda)}\frac{d\tau}{d\lambda}\cr
&=-\sqrt{2(H-\Phi)k_{AB}\frac{dx^A}{d\lambda}\frac{dx^B}{d\lambda}}}
\end{equation}
that acts also as a Lagrangian for timelike orbits of the original dynamical system. This implies that \emph{the timelike orbits are geodesics of the configuration space endowed with the Jacobi metric (\ref{II.3}).}

In local coordinates, the geodesics equations are given by
\begin{equation}\label{II.9}
\frac{d^2x^A}{ds^2}+\conn{A}{B}{C}\frac{dx^B}{ds}\frac{dx^C}{ds}=0,
\end{equation}
where
\begin{equation}\label{II.10}
s(\tau)=\int_{\tau_0}^\tau\,\sqrt{-g_{AB}\frac{dx^A}{d\lambda}\frac{dx^B}{d\lambda}}\,d\lambda
\end{equation}
is the usual affine parameter and 
\begin{equation}\label{II.11}
\conn{A}{B}{C}=\frac{g^{AD}}{2}(g_{DB,C}+g_{DC,B}-g_{BC,D})
\end{equation}
are the Christoffel symbols. As is widely recognized, the local stability of a fiducial geodesic is governed by the Jacobi separation field $J^A$ that evolves according to the Jacobi equations
\begin{equation}\label{II.12}
\frac{\nabla^2J^A}{ds^2}+\riem{A}{B}{C}{D}\frac{dx^B}{ds}J^C\frac{dx^D}{ds}=0,
\end{equation}
where 
\begin{equation}\label{II.13}
\frac{\nabla J^A}{ds}=\frac{dJ^A}{ds}+\conn{A}{B}{C}\frac{dx^B}{ds}J^C
\end{equation}
is the covariant derivative of $J^A$ and 
\begin{equation}\label{II.14}
\riem{A}{B}{C}{D}=\conn{A}{B}{D,C}-\conn{A}{B}{C,D}+\conn{A}{E}{C}\conn{E}{B}{D}-\conn{A}{E}{D}\conn{E}{B}{C}
\end{equation}
are the components of the Riemann curvature tensor. More explicitly, the Jacobi equations are given by
\begin{equation}\label{II.15}
\frac{d^2J^A}{ds^2}+2\conn{A}{B}{C}\frac{dx^B}{ds}\frac{dJ^C}{ds}+\conn{A}{B}{C,D}\frac{dx^B}{ds}\frac{dx^C}{ds}J^D=0.
\end{equation}
This way we have translated the stability problem of orbits in the original mechanical system into geometric language.

Specializing to the case $k_{AB}=\eta_{AB}$, so that $g_{AB}=e^{2\sigma}\eta_{AB}$, where 
\begin{equation}\label{II.16}
\sigma=\frac{\ln\left[2(\Phi-H)\right]}{2}
\end{equation}
and
\begin{equation}\label{II.17}
\left[\eta_{AB}\right]=
\left[\begin{array}{cccccc}-1&0&0&\cdots&0&0\\
0&1&0&\cdots&0&0\\
0&0&1&\cdots&0&0\\
\vdots&\vdots&\vdots&\ddots&\vdots&\vdots\\
0&0&0&\cdots&1&0\\
0&0&0&\cdots&0&1
\end{array}\right]
\end{equation}
is the usual Lorentzian metric tensor, we have
\begin{equation}\label{II.18}
\conn{A}{B}{C}=\krod{A}{B}\sigma_{,C}+\krod{A}{C}\sigma_{,B}-\eta_{BC}\sigma^{,A},
\end{equation}
\begin{equation}\label{II.19}
\conn{A}{B}{C,D}=\krod{A}{B}\sigma_{,CD}+\krod{A}{C}\sigma_{,BD}-\eta_{BC}\sigma^{,A}_{\ \ ,D},
\end{equation}
and
\begin{eqnarray}\label{II.20}
\fl\riem{A}{B}{C}{D}=\krod{A}{D}(\sigma_{,BC}-\sigma_{,B}\sigma_{,C})-\krod{A}{C}(\sigma_{,BD}-\sigma_{,B}\sigma_{,D})+(\krod{A}{D}\eta_{BC}-\krod{A}{C}\eta_{BD})\sigma^{,E}_{\ \ ,E}\cr
+\eta_{BC}(\sigma^{,A}_{\ \ ,D}-\sigma^{,A}\sigma_{,D})-\eta_{BD}(\sigma^{,A}_{\ \ ,C}-\sigma^{,A}\sigma_{,C}).
\end{eqnarray}
So the geodesic equations are given by
\begin{equation}\label{II.21}
\frac{d^2x^A}{ds^2}+2\frac{d\sigma}{ds}\frac{dx^A}{ds}-\sigma^{,A}\eta_{BC}\frac{dx^B}{ds}\frac{dx^C}{ds}=0
\end{equation}
and, upon using the relation $ds/d\tau=e^{2\sigma}$ that follows from (\ref{II.6}), (\ref{II.10}), and (\ref{II.16}), by
\begin{equation}\label{II.22}
\frac{d^2x^A}{d\tau^2}-\sigma^{,A}\eta_{BC}\frac{dx^B}{d\tau}\frac{dx^C}{d\tau}=0
\end{equation}
while parametrized by the original orbit parameter $\tau$. Using (\ref{II.6}) with $\lambda=\tau$ and $k_{AB}=\eta_{AB}$, it can be shown that (\ref{II.22}) is equivalent to
\begin{equation}\label{II.23}
\frac{d^2x^A}{d\tau^2}+\sigma^{,A}e^{2\sigma}=0.
\end{equation}
Since 
\begin{equation}\label{II.24}
\sigma_{,A}e^{2\sigma}=\Phi_{,A}
\end{equation}
by (\ref{II.16}), (\ref{II.23}) is in turn equivalent to
\begin{equation}\label{II.25}
\frac{d^2x^A}{d\tau^2}+\eta^{AB}\Phi_{,B}=0.
\end{equation}
These are simply the Euler's equations associated with the original Lagrangian ${\cal L}$ with $k_{AB}=\eta_{AB}$.

On the other hand, since $e^{2\sigma}\eta_{AB}\left(dx^A/ds\right)\left(dx^B/ds\right)=-1$ by (\ref{II.3}), (\ref{II.10}), and (\ref{II.16}), the Jacobi equations for timelike orbits are given by
\begin{equation}\label{II.26}
\fl\frac{d^2J^A}{ds^2}+2\frac{d\sigma}{ds}\frac{dJ^A}{ds}+2\frac{dx^A}{ds}\frac{d}{ds}\left(\sigma_{,B}J^B\right)-2\sigma^{,A}\eta_{BC}\frac{dx^B}{ds}\frac{dJ^C}{ds}+\sigma^{,A}_{\ \ ,B}J^Be^{-2\sigma}=0.
\end{equation}
Using the relation $ds/d\tau=e^{2\sigma}$, the Jacobi equations are also given by
\begin{equation}\label{II.27}
\fl\frac{d^2J^A}{d\tau^2}+2\left(\frac{dx^A}{d\tau}\sigma_{,B}-\sigma^{,A}\eta_{BC}\frac{dx^C}{d\tau}\right)\frac{dJ^B}{d\tau}
+\left(2\frac{dx^A}{d\tau}\sigma_{,BC}\frac{dx^C}{d\tau}+\sigma^{,A}_{\ \ ,B}e^{2\sigma}\right)J^B=0.
\end{equation}

Denoting the coordinate $x^0$ by $t$, we call $t$ the lab time and an orbit along which $x^A$ remains constant for all $A\ne 0$ a \emph{spatially constant orbit}. On a spatially constant orbit, $\sigma_{,A}$ vanishes identically for all $A\ne 0$ by any of the geodesic equations (\ref{II.21}), (\ref{II.22}), or (\ref{II.23}), and $dt/d\tau=e^\sigma$ by (\ref{II.6}) and (\ref{II.16}). Hence on such an orbit, the Jacobi equations can be reduced to
\begin{equation}\label{II.28}
\frac{d^2J^A}{ds^2}+2\frac{d\sigma}{ds}\frac{dJ^A}{ds}+e^{-2\sigma}\sigma_{,AB}J^B=0,
\end{equation}
\begin{equation}\label{II.29}
\frac{d^2J^A}{d\tau^2}+e^{2\sigma}\sigma_{,AB}J^B=0,
\end{equation}
and
\begin{equation}\label{II.30}
\frac{d^2J^A}{dt^2}+\frac{d\sigma}{dt}\frac{dJ^A}{dt}+\sigma_{,AB}J^B=0
\end{equation}
while being parametrized by the affine parameter $s$, the original orbit parameter $\tau$, and the lab time $t$ respectively.

\section{Dynamics of a Charge in a Generic E.M. Wave with Planar Symmetry}
\label{section III}

We begin this section with the reduction of the interaction of a charged particle and a generic E.M. field with planar symmetry to its longitudinal degrees of freedom. Then we show that the motion of the particle in the direction longitudinal to the field propagation is a natural mechanical system and hence the corresponding orbits are the geodesics in the reduced configuration space with a suitable Jacobi metric.

A generic E.M. with planar symmetry is defined by the vanishing of the longitudinal components of its vector potential and its sole dependence on the longitudinal coordinates:
\begin{equation}\label{III.1}
A_t=A_x=0,\ A_y=A_y(t,x),\ A_z=A_z(t,x).
\end{equation}
Together with the Lorentz equations of motion
\begin{equation}\label{III.2}
m\frac{d^2x^\alpha}{d\tau^2}=q\eta^{\alpha\beta}\left(\frac{\partial A_\gamma}{\partial x^\beta}-\frac{\partial A_\beta}{\partial x^\gamma}\right)\frac{dx^\gamma}{d\tau},
\end{equation}
the vector potential $A_\alpha$ determines completely the dynamics of a particle of mass $m$ and charge $q$ in the E.M. field. As a convention, all Greek letters $\alpha,\beta,\gamma$, and so on, whenever appearing as subscripts or superscripts, always run over the values $0,1,2$, and $3$. We will also use capital and lower-case Latin letters to denote the longitudinal and transverse degrees of freedom respectively.

The solutions of the Lorentz equations are the world lines 
\begin{equation}\label{III.3}
(x^\alpha(\tau))=(x^A(\tau);x^a(\tau))=(t(\tau),x(\tau);y(\tau),z(\tau)),
\end{equation}
each of which is parametrized by the particle's proper time $\tau$. The coordinate $t=x^0$ is the lab time in units of light traveling distance and is related to the conventional time $t_{conv}$ by
\begin{equation}\label{III.4}
t=ct_{conv}.
\end{equation}
As is generally known, the Lorentz equations are the Euler-Lagrange's equations of the Lagrangian
\begin{equation}\label{III.5}
{\cal L}^{(4)}\left(x^\alpha,\frac{dx^\alpha}{d\tau}\right)=\frac{1}{2}\eta_{\beta\gamma}\frac{dx^\beta}{d\tau}\frac{dx^\gamma}{d\tau}+\frac{q}{m}A_\beta(x^\alpha)\frac{dx^\beta}{d\tau}
\end{equation}
with the equivalent Hamiltonian
\begin{equation}\label{III.6}
{\cal H}^{(4)}\left(p_\alpha,x^\alpha\right)=\frac{\eta^{\beta\gamma}\left(p_\beta-\frac{q}{m}A_\beta\right)\left(p_\gamma-\frac{q}{m}A_\gamma\right)}{2},
\end{equation}
where $p_\alpha=\partial{\cal L}^{(4)}/\partial(dx^\alpha/d\tau)=\eta_{\alpha\beta}dx^\beta/d\tau+qA_\alpha/m$ are the generalized momenta. As a consequence of the synchronization of the laboratory clock and a clock comoving with the particle whenever the particle is at rest $\left(dx/d\tau=dy/d\tau=dz/d\tau=0\right)$ in the lab frame, it follows that ${\cal H}^{(4)}=-\frac{1}{2}$.

The planar symmetry of the vector potential induces the cyclic nature of the transverse coordinates $y$ and $z$ with respect to the Lagrangian ${\cal L}^{(4)}$. This leads to the fact that the transverse momenta
\begin{equation}\label{III.7}
P_a=\eta_{ab}\frac{dx^b}{d\tau}+\frac{q}{m}A_a(t,x)
\end{equation}
are integrals of motion besides the Hamiltonian ${\cal H}^{(4)}$. Thanks to these integrals of motion, we can reduce the number of degrees of freedom of the particle dynamics using the classical Routh's procedure. To this end, we solve for
\begin{equation}\label{III.8}
\frac{dx^a}{d\tau}=\left(P_b-\frac{qA_b}{m}\right)\eta^{ba}
\end{equation}
from (\ref{III.7}) and substitute them into
\begin{equation}\label{III.9}
\eqalign{{\cal L}^{(2)}\left(x^A,\frac{dx^A}{d\tau}\right)&={\cal L}^{(4)}-\frac{\partial{\cal L}^{(4)}}{\partial(dx^A/d\tau)}\frac{dx^a}{d\tau}\cr
&=\frac{\eta_{BC}}{2}\frac{dx^B}{d\tau}\frac{dx^C}{d\tau}-\Phi(t,x),}
\end{equation}
where the scalar potential is given by
\begin{equation}\label{III.10}
\Phi(t,x)=\frac{\eta^{ab}\left(P_a-\frac{q}{m}A_a\right)\left(P_b-\frac{q}{m}A_b\right)}{2},
\end{equation}
to get a Lagrangian ${\cal L}^{(2)}$ for the motion of the particle in the direction longitudinal to the field propagation.

The Routh's reduction indicates that \emph{the dynamics of the E.M. field-accelerated particle is controlled entirely by the equation
\begin{equation}\label{III.11}
\frac{d^2x^A}{d\tau^2}=-\eta^{AB}\Phi_{,B},
\end{equation}
which are the Euler-Lagrange's equations for the reduced Lagrangian ${\cal L}^{(2)}$.} By contrast, \emph{subsequent to the launching of the particle, the transverse degrees of freedom have no active effect on the particle dynamics in the $tx$-plane and are governed by the longitudinal dynamics through (\ref{III.7});} that is, the transverse degrees of freedom enter the longitudinal dynamics of the particle only as parameters by means of the initial data.

On the other hand, upon comparing (\ref{III.9}) to (\ref{II.1}), we note that more significant is the fact that the longitudinal dynamics is a natural mechanical system with the Hamiltonian
\begin{equation}\label{III.12}
{\cal H}^{(2)}(p_A,x^A)=\frac{\eta^{BC}p_Bp_C}{2}+\Phi(t,x),
\end{equation}
where $p_A=\partial{\cal L}^{(2)}/\partial(dx^A/d\tau)=\eta_{AB}dx^B/d\tau$ are the generalized momenta. Also, along any physical orbit, we have
\begin{equation}\label{III.13}
{\cal H}^{(2)}={\cal H}^{(4)}=-\frac{1}{2}
\end{equation}
by (\ref{III.6}) and (\ref{III.10}). In consequence, \emph{the longitudinal dynamics can be and will be identified with the geodesics in the $tx$-plane endowed with the metric tensor
\begin{equation}\label{III.14}
{^{(2)}}g_{AB}=\left(1+2\Phi\right)\eta_{AB}=e^{2\sigma}\eta_{AB},
\end{equation}}
\noindent
as is given in the more general case by (\ref{II.3}). 

Being defined on a two-dimensional manifold, the geodesic equations have the simple form
\numparts
\begin{eqnarray}\label{III.15}
\frac{d^2u}{ds^2}+2\sigma_{,u}\left(\frac{du}{ds}\right)^2&=0,\label{III.15a}\\
\frac{d^2v}{ds^2}+2\sigma_{,v}\left(\frac{dv}{ds}\right)^2&=0\label{III.15b}
\end{eqnarray}
\endnumparts
while parametrized by the affine parameter $s$, where
\begin{equation}\label{III.16}
u=t-x\mbox{ and }v=t+x,
\end{equation}
are the null coordinates. The components of the Riemann curvature tensor can be reduced to
\begin{equation}\label{III.17}
{^{(2)}}\riem{A}{B}{C}{D}=e^{2\sigma}K\left(\krod{A}{C}\eta_{BD}-\krod{A}{D}\eta_{BC}\right),
\end{equation}
where
\begin{equation}\label{III.18}
K=-e^{-2\sigma}\sigma^{,A}_{\ \ ,A}
\end{equation}
is the Gaussian curvature. Thus the Jacobi equations (\ref{II.12}) take the form
\begin{equation} \label{III.19}
\frac{\nabla^2J^A}{ds^2}+e^{2\sigma}K\left(\krod{A}{C}\eta_{BD}-\krod{A}{D}\eta_{BC}\right)\frac{dx^B}{ds}J^C\frac{dx^D}{ds}=0.
\end{equation}
This way \emph{one sees that the stability question of the orbit of a charged particle in response to an E.M. field with planar symmetry is encapsulated into the product of two geometrical quantities, i.e., the conformal factor $e^{2\sigma}$ and the Gaussian curvature $K$.}
\section{Dynamics of a Charge in a Curvature-Free Wave Field with Planar Symmetry}
\label{section IV}

It is clear from (\ref{III.18}) that the $tx$-plane endowed with the metric ${^{(2)}}g_{AB}$ is flat if and only if
$\sigma(u,v)=U(u)+V(v)$, where $U$ and $V$ are smooth functions of one variable. Under this circumstance, initially parallel geodesics preserve their separations \cite[$\S$11.5]{misner1973}. In addition, with the vanishing of the Gaussian curvature $K$, the geodesic equations (\ref{III.15a}) and (\ref{III.15b}) decouple from each other and are equivalent to
\begin{equation}\label{IV.1}
\frac{d}{ds}\left(e^{2U(u)}\frac{du}{ds}\right)=\frac{d}{ds}\left(e^{2V(v)}\frac{dv}{ds}\right)=0.
\end{equation}
Consequently, one has
\begin{equation}\label{IV.2}
\int\,e^{2U(u)}\,du=\mbox{const.}\int\,ds\mbox{ and }\int\,e^{2V(v)}\,dv=\mbox{const.}\int\,ds.
\end{equation}
In summary, this shows that \emph{the vanishing of the Gaussian curvature $K$ implies the integrability by quadratures of the particle motion.}

A curvature-free instance of physical interest is readily furnished by an elliptically polarized plane-wave field propagating along the $x$-axis. The components of the vector potential can be represented by 
\begin{equation}\label{IV.3}
A_t=A_x=0,\ A_y(u)=\frac{{\cal E}\delta}{\omega}\cos\omega u,\ A_z(u)=\frac{{\cal E}\sqrt{1-\delta^2}}{\omega}\sin\omega u,
\end{equation}
where ${\cal E}$ and $\omega$ are the field amplitude and frequency respectively, while $\delta$ is a polarization parameter such that $\delta=0,\pm 1$ for a linearly polarized wave and $\delta=\pm 1/\sqrt{2}$ for a circular wave. It can be shown that the equations of motion for a particle in such a field can be solved exactly \cite{bardsley1989,gibbon2005,hartemann1998,sarachik1970}. Since the vector potential depends only on $u=t-x$, it is readily seen from (\ref{II.16}), (\ref{III.10}), and (\ref{III.18}) that the induced $tx$-plane is flat. Thus the widely known integrability of the particle motion is indeed a consequence of the vanishing of the Gaussian curvature $K$. Furthermore, the tranverse motion can be obtained by integrating (\ref{III.7}), while the longitudinal dynamics of the particle can be determined with the aid of (\ref{IV.2}) by integrating 
\begin{equation}\label{IV.4}
\fl\int\,\left[1+\left(P_y+\eta\delta\cos\omega u\right)^2+\left(P_z+\eta\sqrt{1-\delta^2}\sin\omega u\right)^2\right]\,du\\=\mbox{const.}\int\,dv,
\end{equation}
where we recall $\eta=q{\cal E}/m\omega$ is the dimensionless impulse factor of the interaction.
\section{Dynamics of a Charge in a Standing Wave Field with Planar Symmetry}
\label{section V}

In general, the Gaussian curvature $K$ as determined from (\ref{III.18}) is nonvanishing. Of particular interest is the motion of a charged particle driven by the E.M. field of a standing wave \cite{bauer1995,kaplan2005,pokrovsky2005} with planar symmetry and polarized linearly along the $z$-axis. The components of the vector potential are
\begin{equation}\label{V.0.1}
A_t=A_x=A_y=0,\ A_z(t,x)=\frac{{\cal E}}{\omega}\sin\omega t\sin\omega x,
\end{equation}
where ${\cal E}$ and $\omega$ are the field amplitude and frequency respectively. In this case, the scalar potential for the longitudinal dynamics is given by
\begin{equation}\label{V.0.2}
\Phi(t,x)=\frac{1}{2}\left[P_y^2+(P_z-\eta\sin\omega t\sin\omega x)^2\right].
\end{equation}
Here we recall that
\begin{equation}\label{V.0.3}
\eta=\frac{q{\cal E}}{m\omega}
\end{equation}
is the dimensionless impulse factor of the interaction. The different numerical subintervals within which the value of $\eta$ lies, as we will see, play a significant role in the local stability of the physical orbits of the particle.

By (\ref{III.11}), the longitudinal dynamics is governed by the equations
\numparts
\begin{eqnarray}\label{V.0.4}
\frac{d^2t}{d\tau^2}&=-\frac{\eta}{\omega}(P_z-\eta\sin\omega t\sin\omega x)\cos\omega t\sin\omega x,\label{V.0.4a}\\
\frac{d^2x}{d\tau^2}&=\frac{\eta}{\omega}(P_z-\eta\sin\omega t\sin\omega x)\sin\omega t\cos\omega x.\label{V.0.4b}
\end{eqnarray}
\endnumparts
If, in addition, we stipulate that $P_z=0$, then there are two classes of spatially constant orbits. One class of these spatially constant orbits are those along which $\sin\omega x=0$ and whose $x$-coordinates are
\begin{equation}\label{V.0.5}
x(\tau)=\pm\frac{n\pi}{\omega},\ n=0,1,2,\cdots,
\end{equation}
and another class of spatially constant orbits are those along which $\cos\omega x=0$ and whose $x$-coordinates are
\begin{equation}\label{V.0.6}
x(\tau)=\pm\left(n+\frac{1}{2}\right)\frac{\pi}{\omega},\ n=0,1,2,\cdots.
\end{equation}
We will use the first class of spatially constant orbits to discuss their local stability characterized in terms of the numerical subintervals of the impulse factor $\eta$.
\subsection{Dynamics about the Orbits $x=\pm n\pi/\omega$}
\label{section V.A}

By setting $\sigma=\frac{1}{2}\ln(2\Phi+1)$ in accordance with (\ref{II.16}) and (\ref{III.13}), one finds
\numparts
\begin{eqnarray}\label{V.A.1}
\sigma_{,t}&=\frac{\eta^2\omega\cos\omega t\sin\omega t\sin^2\omega x}{1+P_y^2+\eta^2\sin^2\omega t\sin^2\omega x},\label{V.A.1a}\\
\sigma_{,x}&=\frac{\eta^2\omega\sin^2\omega t\cos\omega x\sin\omega x}{1+P_y^2+\eta^2\sin^2\omega t\sin^2\omega x},\label{V.A.1b}\\
\sigma_{,tt}&=(\eta\omega)^2\frac{(1+P_y^2)\cos^2\omega t\sin^2\omega x-\eta^2\sin^2\omega t\sin^4\omega x}{(1+P_y^2+\eta^2\sin^2\omega t\sin^2\omega x)^2},\label{V.A.1c}\\
\sigma_{,tx}&=2(\eta\omega)^2\frac{(1+P_y^2)\cos\omega t\sin\omega t\cos\omega x\sin\omega x}{(1+P_y^2+\eta^2\sin^2\omega t\sin^2\omega x)^2},\label{V.A.1d}\\
\sigma_{,xx}&=(\eta\omega)^2\frac{(1+P_y^2)\sin^2\omega t\cos^2\omega x-\eta^2\sin^4\omega t\sin^2\omega x}{(1+P_y^2+\eta^2\sin^2\omega t\sin^2\omega x)^2}.\label{V.A.1e}
\end{eqnarray}
\endnumparts
In particular, along the spatially constant orbits $x=\pm n\pi/\omega$, $n=0,1,2,\cdots$, one has
\begin{equation}\label{V.A.2}
\eqalign{
\sigma_{,t}=\sigma_{,x}=0\mbox{ and }\sigma_{,tt}=\sigma_{,tx}=0,\label{V.A.2a}\\
\sigma_{,xx}(t,x)=\frac{(\eta\omega\sin\omega t)^2}{1+P_y^2}.\label{V.A.2b}}
\end{equation}
It follows from these equations and (\ref{II.30}) that the Jacobi equations associated with the spatially constant orbits $x=\pm n\pi/\omega$, $n=0,1,2,\cdots$, are
\numparts
\begin{eqnarray}\label{V.A.3}
\frac{d^2J^t}{dt^2}&=0,\label{V.A.3a}\\
\frac{d^2J^x}{dt^2}+\frac{(\eta\omega\sin\omega t)^2}{1+P_y^2}J^x&=0\label{V.A.3b}
\end{eqnarray}
\endnumparts
while being parametrized by the lab time $t$. Due to their simplicity in form, they readily admit further analysis to address the dynamical properties of neighboring orbits of the spatially constant orbits $x=\pm n\pi/\omega$, $n=0,1,2,\cdots$. By a neighboring orbit of a spatially constant orbit we mean one which begins in a sufficiently small neighborhood centered around the initial datum for the particular spatially constant orbit in the phase space of initial data.

Without restricting generality, let $P_y=0$ and let the fiducial orbit be the spatially constant orbit $x=0$. On rescaling the lab time using the optical cycle (i.e., $2\pi/\omega$ units of lab time), (\ref{V.A.3b}) becomes
\begin{equation}\label{V.A.4}
\frac{d^2J^x}{dT^2}+(2\pi\eta\sin 2\pi T)^2J^x=0,
\end{equation}
where
\begin{equation}\label{V.A.5}
T=\frac{\omega t}{2\pi}
\end{equation}
is the dimensionless time in units of optical cycles. Since $(2\pi\eta)^2\int_0^\infty\,\sin^22\pi T\,dT=\infty$, (\ref{V.A.3b}) is oscillatory in the sense that all its solutions have arbitrarily large zeros \cite[Theorem 2.4.1]{agarwal2002}. In other words, the fiducial orbit $x=0$ has infinitely many conjugate points. Hence we expect that neighboring orbits intersect the fiducial orbit infinitely often, in particular, whenever the corresponding solutions of (\ref{V.A.4}) are bounded. The distribution of these conjugate points may therefore serve as a physically identifiable diagnostic property of the dynamics of the neighboring orbits of the orbit $x=0$ as estimated by (\ref{V.A.4}) as compared to (\ref{V.0.4a}) and (\ref{V.0.4b})

More pertinent to the stability question of the spatially constant orbit $x=0$ is the fact that (\ref{V.A.4}) is a Hill equation, so that the parametric dependence on the impulse factor $\eta$ of the boundedness of its solutions can be deduced from standard Floquet theory as follows. Let $J_1^x$ and $J_2^x$ be the solutions of (\ref{V.A.4}) such that 
\begin{equation}\label{V.A.6}
J_1^x=\frac{dJ_2^x}{dT}=1\mbox{ and }\frac{dJ_1^x}{dT}=J_2^x=0\mbox{ at }T=0.
\end{equation}
Following \cite{yakubovich1975-2}, the characteristic function $\varphi(\eta)$ of (\ref{V.A.4}) is defined as half the trace of its fundamental matrix at a period, i.e.,
\begin{equation}\label{V.A.7}
\varphi(\eta)=\left.\frac{1}{2}\tr\left[\begin{array}{cc}J^x_1&J^x_2\\dJ^x_1/dT&dJ^x_2/dT\end{array}\right]\right|_{T=\frac{1}{2}}.
\end{equation}
Then it is known that all solutions of (\ref{V.A.4}) are bounded if and only if $|\varphi(\eta)|<1$ \cite[Sec. 2.4]{chicone2006}. Since the characteristic function $\varphi(\eta)$ depends generally on the value of $\eta$, so does the boundedness of solutions of (\ref{V.A.4}) and hence the stability of the spatially constant orbit $x=0$. Viewing the impulse factor $\eta$ as a parameter of the system, we call any value of $\eta$ for which $x=0$ becomes unstable a resonant value and say that there is a parametric resonance under such a circumstance.

In fact, $|\varphi(\eta)|-1$ has arbitrarily large zeros on the interval $(0,\infty)$. Thus \emph{there exist infinitely many disjoint open subintervals  $(\eta_k^+,\eta_{k+1}^-)\subset(0,\infty)$, $k=0,1,2,\cdots$, ordered according to their left endpoints, in which $|\varphi(\eta)|<1$ \cite[Ch. VII, $\S$1.4, $\S$1.5]{yakubovich1975-2}  and for values of $\eta$ lying within these intervals, the spatially constant orbit $x=0$ is stable.} These subintervals are called the stability zones; the complementary closed subintervals, within the interior of each of which $|\varphi(\eta)|>1$, are called the instability zones.  As an additional remark, it follows from a classical result \cite{borg1949} that if
\begin{equation}\label{V.A.8}
\frac{\eta^2}{4}=\eta^2\int_0^{1/2}\,\sin^22\pi T\,dT\le\frac{2}{\pi^2},
\end{equation}
then all solutions of (\ref{V.A.4}) are bounded. Hence the first stability zone must be of the form $(0,\eta_1^-)$ for some $\eta_1^-\ge\frac{2\sqrt{2}}{\pi}\approx 0.900316$. Numerically, $\eta_1^-\approx 1.147179$. In consequence, \emph{the orbit $x=0$ is stable in a sufficiently low impulsive E.M. field ($\eta\ll 1$).}
\subsection{A Mathematically Precise Averaging Principle for the Jacobi Equation}
\label{section V.B}

In a sufficiently low impulsive E.M. field ($\eta\ll 1$), the dynamics of the neighboring orbits of the spatially constant orbit $x=0$ can be analyzed using the method of averaging. To this end, we note that (\ref{V.A.4}) is equivalent to the linear periodic system
\begin{equation}\label{V.B.1}
\eqalign{
\frac{d\boldsymbol{J}}{dT}&=2\pi\eta A(T)\boldsymbol{J},\label{V.B.1a}\cr
&\fl\mbox{where}\cr
\boldsymbol{J}&=\left[\begin{array}{c}J_1\\ J_2\end{array}\right]=\left[\begin{array}{cc}J^x\\ (2\pi\eta)^{-1}dJ^x/dT\end{array}\right]\label{V.B.1b}\cr
&\fl\mbox{and the coefficient matrix}\cr
A(T)&=\left[\begin{array}{cc}0&1\\-\sin^22\pi T&0\end{array}\right]\label{V.B.1c}}
\end{equation}
has period half an optical cycle. Associated with (\ref{V.B.1}) is the averaged system \cite{hoppensteadt2000}
\begin{equation}\label{V.B.2}
\eqalign{
\frac{d\overline{\boldsymbol{J}}}{dT}&=2\pi\eta\overline{A}\overline{\boldsymbol{J}},\label{V.B.2a}\cr
&\fl\mbox{where}\cr
\overline{A}&=\left[\begin{array}{cc}0&1\\-\frac{1}{2}&0\end{array}\right]\label{V.B.2b}}
\end{equation}
is the average of $A$ over one of its own cycles or over an optical cycle. The solution of the averaged system, i.e., (\ref{V.B.2}), with the initial condition $\overline{\boldsymbol{J}}(0)=\left[\begin{array}{c}\overline{J}_{10}\\\overline{J}_{20}\end{array}\right]$ is readily obtained as
\begin{equation}\label{V.B.3}
\eqalign{
\overline{\boldsymbol{J}}(T)&=e^{2\pi\eta\overline{A}T}\overline{\boldsymbol{J}}(0)\cr
&=\left[\begin{array}{cc}\cos\sqrt{2}\pi\eta T&\sqrt{2}\sin\sqrt{2}\pi\eta T\\-\frac{1}{\sqrt{2}}\sin\sqrt{2}\pi\eta T&\cos\sqrt{2}\pi\eta T\end{array}\right]\left[\begin{array}{c}\overline{J}_{10}\\\overline{J}_{20}\end{array}\right],}
\end{equation}
showing that the averaged system exhibits simple harmonic motions with the frequency $\eta/\sqrt{2}$ oscillations per optical cycle.  To consider how good the averaged system (\ref{V.B.2}) approximates the original system (\ref{V.B.1}), let
\begin{equation}\label{V.B.4}
\eqalign{
\left[\begin{array}{c}E_{1}\\ E_{2}\end{array}\right]&=\boldsymbol{E}=\boldsymbol{J}-\overline{\boldsymbol{J}}=\left[\begin{array}{c}J_{1}-\overline{J}_{1}\\ J_{2}-\overline{J}_{2}\end{array}\right]\label{V.B.4a}\cr
&\fl\mbox{be the error in estimating $\boldsymbol{J}$ using $\overline{\boldsymbol{J}}$ with the initial error}\cr
\left[\begin{array}{c}E_{10}\\ E_{20}\end{array}\right]&=\boldsymbol{E}(0)=\boldsymbol{J}(0)-\overline{\boldsymbol{J}}(0)=\left[\begin{array}{c}J_{10}-\overline{J}_{10}\\ J_{20}-\overline{J}_{20}\end{array}\right].\label{V.B.4b}}
\end{equation}
Calculating $\boldsymbol{E}(T)$ from (\ref{V.B.1}), (\ref{V.B.2}), and (\ref{V.B.4}) yields
\begin{equation}\label{V.B.5}
\eqalign{\boldsymbol{E}(T)-\boldsymbol{E}(0)
&=2\pi\eta\int_0^T\,\left[A(S)(\boldsymbol{E}(S)-\boldsymbol{E}(0))\right.\cr
&\qquad\left.+(A(S)-\overline{A})\overline{\boldsymbol{J}}(S)+A(S)\boldsymbol{E}(0)\right]\,dS.}
\end{equation}
In consequence, one finds
\begin{equation}\label{V.B.6}
\fl\eqalign{
&\quad\left\|\boldsymbol{E}(T)-\boldsymbol{E}(0)\right\|\cr
&\le2\pi\eta\int_0^T\,\left[\left\|A(S)(\boldsymbol{E}(S)-\boldsymbol{E}(0))\right\|+\left\|\left(A(S)-\overline{A}\right)\overline{\boldsymbol{J}}(S)\right\|+\left\|A(S)\boldsymbol{E}(0)\right\|\right]\,dS\cr
&\le2\pi\eta\int_0^T\,\sqrt{(E_1(S)-E_{10})^2\sin^42\pi S+(E_2(S)-E_{20})^2}\,dS\cr
&\qquad+\pi\eta\int_0^T\,\left[\left|\cos4\pi S\left(\overline{J}_{10}\cos\sqrt{2}\pi\eta S+\overline{J}_{20}\sqrt{2}\sin\sqrt{2}\pi\eta S\right)\right|\right.\cr
&\qquad\qquad\left.+2\sqrt{E_{10}^2\sin^42\pi S+E_{20}^2}\right]\,dS\cr
&\le2\pi\eta\int_0^T\,\left\|(\boldsymbol{E}(S)-\boldsymbol{E}(0)\right\|\,dS+\pi\eta\left(\left|\overline{J}_{10}\right|+\sqrt{2}\left|\overline{J}_{20}\right|+2\left\|\boldsymbol{E}(0)\right\|\right)T.}
\end{equation}
Then an application of Gronwall's lemma \cite[Lemma 1.3.3]{sanders2007} yields
\begin{equation}\label{V.B.7}
\left\|\boldsymbol{E}(T)-\boldsymbol{E}(0)\right\|\le\frac{\left|\overline{J}_{10}\right|+\sqrt{2}\left|\overline{J}_{20}\right|+2\left\|\boldsymbol{E}(0)\right\|}{2}\left(e^{2\pi\eta T}-1\right).
\end{equation}
This equation implies that if $\boldsymbol{J}$ and $\overline{\boldsymbol{J}}$ agree initially, then
\begin{equation}\label{V.B.8}
\boldsymbol{J}(T)=\overline{\boldsymbol{J}}(T)+\Or(\varepsilon)
\end{equation}
uniformly for $0\le T\le\ln(1+\varepsilon)/2\pi\eta$ in the sense that 
\begin{equation}\label{V.B.9}
\left\|\boldsymbol{J}(T)-\overline{\boldsymbol{J}}(T)\right\|\le\frac{\left|J_{10}\right|+\sqrt{2}\left|J_{20}\right|}{2}\varepsilon
\end{equation}
for any $\varepsilon>0$. Specifically,
\begin{equation}\label{V.B.10}
J^x(T)=\overline{J^x}(T)+\Or(\varepsilon),
\end{equation}
where
\begin{equation}\label{V.B.11}
\overline{J^x}(T)=J_x(0)\cos\sqrt{2}\pi\eta T+\frac{1}{\sqrt{2}\pi\eta}\left.\frac{dJ^x}{dT}\right|_{T=0}\sin\sqrt{2}\pi\eta T
\end{equation}
is the average value of $J^x=J_{10}$ as obtained from the averaged system (\ref{V.B.2}), i.e., $\overline{J^x}=\overline{J}_{10}$, and the initial data for $J^x$ and $\overline{J^x}$ have been identified. Hence \emph{in a sufficiently low impulsive E.M. field ($\eta\ll 1$), a charged particle that is launched from a neighboring orbit of the spatially constant orbit $x=0$  exhibits an almost simple harmonic motion and oscillates about $x=0$ with a frequency close to $\eta/\sqrt{2}$ oscillations per optical cycle.}
\subsection{The Landau Decomposition of the Jacobi Equation}
\label{section V.C}

We can reformulate the Jacobi equation, i.e., (\ref{V.A.4}), as
\begin{equation}\label{V.C.1}
\frac{d^2J^x}{dT^2}=2\pi^2\eta^2(-1+\cos 4\pi T)J^x.
\end{equation}
Thus the dynamics of the Jacobi field $J^x$ can be viewed as the motion of a particle subject to a time-independent potential
\begin{equation}\label{V.C.2}
U(J^x)=(\pi\eta J^x)^2
\end{equation}
and a force
\begin{equation}\label{V.C.3}
f(J^x,T)=2\pi^2\eta^2\cos 4\pi T\cdot J^x
\end{equation}
which varies in time with the frequency $2$ oscillations per optical cycle. The frequency of $f$ is considered high in the sense that it is of higher order of magnitude of the frequency of the motion of the particle in the absence of the force field $f$, i.e., $2\gg\eta/\sqrt{2}$ in this case. Note that this is equivalent to requiring that the E.M. field is of low impulse, i.e., $\eta\ll 2\sqrt{2}$. It is a common practice in plasma physics to decompose the original motion, i.e., the Jacobi field $J^x$ in our
case, into a secular component $X$ describing the \emph{oscillation center orbit} and an \emph{oscillatory component} $\xi$ that is nearly periodic \cite{gibbon2005,landau1981,nicholson1983,schmidt1979}:
\begin{equation}\label{V.C.4}
\eqalign{
J^x(T)&=X(T)+\xi(T)\label{V.C.4a}\cr
&\fl\mbox{with}\cr
\frac{d^2X}{dT^2}&=-2\pi^2\eta^2(X+2\sin^22\pi T\cdot\xi),\label{V.C.4b}\cr
\frac{d^2\xi}{dT^2}&=2\pi^2\eta^2\cos4\pi T\cdot X.\label{V.C.4c}}
\end{equation}
We call the decomposition $(X,\xi)$ a \emph{Landau decomposition}. This is a non-unique representation of $J^x$ which depends on a choice of $X$ and $\xi$ satisfying the last two coupled equtions of (\ref{V.C.4}).

Of particular interest is the following exact representation. It is obtained by applying the initial conditions $J^x(0)$ and $\left.dJ^x/dT\right|_{T=0}$ to $X$ and $dX/dT$ and having the oscillatory component $\xi$ and its velocity $d\xi/dT$ vanish initially. This gives rise to the \emph{compatibility condition}
\begin{equation}\label{V.C.5}
\eqalign{
X(0)&=J^x(0),\ \left.\frac{dX}{dT}\right|_{T=0}=\left.\frac{dJ^x}{dT}\right|_{T=0}\label{V.C.5a}\cr
&\fl\mbox{and}\cr
\xi(0)&=0,\ \left.\frac{d\xi}{dT}\right|_{T=0}=0.\label{V.C.5b}}
\end{equation}
They determine an exact solution to the last two coupled equations of (\ref{V.C.4}) and thus constitute the exact representation of the Landau decomposition $(X,\xi)$ expressed by (\ref{V.C.4}). \emph{With the compatibility condition, each Jacobi field $J^x$ is uniquely determined by a Landau decomposition and vice verse.}
\subsection{The Averaging of the Landau Decomposition}
\label{section V.D}

On the assumption that the E.M. field is of low impulse, the Laudau decomposition (\ref{V.C.4}) can be analyzed using the method of averaging as in the case of (\ref{V.A.4}). Together with the compatibility condition (\ref{V.C.5}), we will show that the Jacobi field $J^x$ and the oscillation center $X$ agree on average in a mathematically precise sense, thereby quantifying the smallness of the amplitude of oscillatory component $\xi$ as compared to that of $X$. Moreover, the result of the calculation is a preliminary in yielding a mathematically solid meaning to the ponderomotive approximation of the oscillation center $X$ in a sufficiently low impulsive E.M. field ($\eta\ll 1$). This will be discussed in more detail in the next section.

To initiate the averaging process on the Landau decomposition (\ref{V.C.4}), we note the equivalence of the last two equations of (\ref{V.C.4}) to the linear periodic system
\begin{equation}\label{V.D.1}
\eqalign{
\frac{d\boldsymbol{K}}{dT}&=\sqrt{2}\pi\eta B(T)\boldsymbol{K},\label{V.D.1a}\cr
&\fl\mbox{where}\cr
\boldsymbol{K}&=\left[\begin{array}{c}K_1\\ K_2\\ K_3\\ K_4\end{array}\right]=\left[\begin{array}{c}X\\ \left(\sqrt{2}\pi\eta\right)^{-1}dX/dT\\ \xi\\ \left(\sqrt{2}\pi\eta\right)^{-1}d\xi/dT\end{array}\right]\label{V.D.1b}\cr
&\fl\mbox{and the coefficient matrix}\cr
B(T)&=\left[\begin{array}{cccc}0&1&0&0\\
-1&0&-2\sin^22\pi T&0\\
0&0&0&1\\
\cos4\pi T&0&0&0\end{array}\right]\label{V.D.1c}}
\end{equation}
has period half an optical cycle. The averaged system corresponding to (\ref{V.D.1}) is \cite{hoppensteadt2000}
\begin{equation}\label{V.D.2}
\eqalign{
\frac{d\overline{\boldsymbol{K}}}{dT}&=\sqrt{2}\pi\eta\overline{B}\overline{\boldsymbol{K}},\label{V.D.2a}\cr
&\fl\mbox{where}\cr
\overline{B}&=\left[\begin{array}{cccc}
0&1&0&0\\
-1&0&-1&0\\
0&0&0&1\\
0&0&0&0\end{array}\right]\label{V.D.2b}}
\end{equation}
is the average of $B$ over one optical cycle. The solution of the averaged system with the initial condition $\overline{\boldsymbol{K}}(0)=\left[\begin{array}{c}\overline{K}_{10}\\\overline{K}_{20}\\\overline{K}_{30}\\ \overline{K}_{40}\end{array}\right]$ is given by
\begin{equation}\label{V.D.3}
\fl\eqalign{
&\quad\overline{\boldsymbol{K}}(T)\cr
&=e^{\sqrt{2}\pi\eta\overline{B}T}\overline{\boldsymbol{K}}(0)\cr
&=\left[\begin{array}{cccc}
\cos\sqrt{2}\pi\eta T&\sin\sqrt{2}\pi\eta T&\cos\sqrt{2}\pi\eta T-1&\sin\sqrt{2}\pi\eta T-\sqrt{2}\pi\eta T\\
-\sin\sqrt{2}\pi\eta T&\cos\sqrt{2}\pi\eta T&-\sin\sqrt{2}\pi\eta T&\cos\sqrt{2}\pi\eta T-1\\
0&0&1&\sqrt{2}\pi\eta T\\
0&0&0&1
\end{array}\right]\cr
&\qquad\times\left[\begin{array}{c}\overline{K}_{10}\\\overline{K}_{20}\\\overline{K}_{30}\\ \overline{K}_{40}\end{array}\right].}
\end{equation}
Note that $\overline{K}_{30}(T)=\overline{K}_{30}+\sqrt{2}\pi\eta\overline{K}_{40}T$. In order to be consistent with the stipulation that the average of $\xi$ vanishes over each optical cycle, we necessitate 
\begin{equation}\label{V.D.4}
\overline{K}_{30}=\overline{K}_{40}=0.
\end{equation}
 It follows from this that
\begin{equation}\label{V.D.5}
\overline{K}_3(T)=\overline{K}_4(T)=0
\end{equation}
identically. This renders $\overline{K}$ purely oscillatory and thus gives a heuristic justification for the compatibility condition (\ref{V.C.5}). With this condition on any solution of the averaged system (\ref{V.D.2}), we analyze how good the averaged system approximates the original system (\ref{V.D.1}) by considering the error 
\begin{equation}\label{V.D.6}
\eqalign{
\left[\begin{array}{c}F_1\\ F_2\\ F_3\\ F_4\end{array}\right]&=\boldsymbol{F}=\boldsymbol{K}-\overline{\boldsymbol{K}}=
\left[\begin{array}{c}K_1-\overline{K}_1\\ K_2-\overline{K}_2\\ K_3\\ K_4\end{array}\right]\label{V.D.6a}\cr
&\fl\mbox{with the initial error}\cr
\left[\begin{array}{c}F_{10}\\ F_{20}\\ F_{30}\\ F_{40}\end{array}\right]&=\boldsymbol{F}(0)=\boldsymbol{K}(0)-\overline{\boldsymbol{K}}(0)=
\left[\begin{array}{c}K_{10}-\overline{K}_{10}\\ K_{20}-\overline{K}_{20}\\ 0\\ 0\end{array}\right],}\label{V.D.6b}
\end{equation}
where we have set $K_3(0)=K_4(0)=0$ in view of the compatibility condition (\ref{V.C.5}).  As for (\ref{V.B.5}), one finds
\begin{equation}\label{V.D.7}
\eqalign{
\boldsymbol{F}(T)-\boldsymbol{F}(0)&=\sqrt{2}\pi\eta\int_0^T\,\left[B(S)(\boldsymbol{F}(S)-\boldsymbol{F}(0))\right.\cr
&\qquad\left.+(B(S)-\overline{B})\overline{\boldsymbol{K}}(S)+A(S)\boldsymbol{F}(0)\right]\,dS.}
\end{equation}
In view of this, one has
\begin{equation}\label{V.D.8}
\fl\eqalign{
&\quad\left\|\boldsymbol{F}(T)-\boldsymbol{F}(0)\right\|\cr
&\le\sqrt{2}\pi\eta\int_0^T\,\left[\left\|B(S)(\boldsymbol{F}(S)-\boldsymbol{F}(0))\right\|+\left\|\left(B(S)-\overline{B}\right)\overline{\boldsymbol{K}}(S)\right\|+\left\|B(S)\boldsymbol{F}(0)\right\|\right]\,dS\cr
&\le\sqrt{2}\pi\eta\int_0^T\,\left[(F_{1}(S)-F_{10})^2(1+\cos^24\pi S)+(F_2(S)-F_{20})^2\right.\cr
&\qquad\left.+F_3(S)^2\sin^42\pi S+F_4(S)^2+2(F_1(S)-F_{10})F_3(S)\sin^22\pi S\right]^{1/2}\,dS\cr
&\quad\qquad+\sqrt{2}\pi\eta\int_0^T\,\left\{\left|\cos4\pi S\left(\overline{K}_{10}\cos\sqrt{2}\pi\eta S+\overline{K}_{20}\sin\sqrt{2}\pi\eta S\right)\right|\right.\cr
&\quad\quad\qquad\left.+\left[F_{10}^2(1+\cos^24\pi S)+F_{20}^2\right]^{1/2}\right\}\,dS\cr
&\le2\sqrt{2}\pi\eta\int_0^T\,\left\|\boldsymbol{F}(S)-\boldsymbol{F}(0)\right\|\,dS+\sqrt{2}\pi\eta\left(\left|\overline{K}_{10}\right|+\left|\overline{K}_{20}\right|+\left\|\boldsymbol{F}(0)\right\|\right)T.}
\end{equation}
Hence an application of Gronwall's lemma \cite[Lemma 1.3.3]{sanders2007} yields
\begin{equation}\label{V.D.9}
\left\|\boldsymbol{F}(T)-\boldsymbol{F}(0)\right\|\le\frac{\left|\overline{K}_{10}\right|+\left|\overline{K}_{20}\right|+\left\|\boldsymbol{F}(0)\right\|}{2}\left(e^{2\sqrt{2}\pi\eta T}-1\right).
\end{equation}
This equation indicates that if $\boldsymbol{K}$ and $\overline{\boldsymbol{K}}$ agree initially, then
\begin{equation}\label{V.D.10}
\boldsymbol{K}(T)=\overline{\boldsymbol{K}}(T)+\Or(\varepsilon)
\end{equation}
uniformly for $0\le T\le\ln(1+\varepsilon)/2\sqrt{2}\pi\eta$ in the sense that
\begin{equation}\label{V.D.11}
\left\|\boldsymbol{K}(T)-\overline{\boldsymbol{K}}(T)\right\|\le\frac{|K_{10}|+|K_{20}|}{2}\varepsilon
\end{equation}
for any $\varepsilon>0$. In particular, one concludes that
\begin{equation}\label{V.D.12}
X(T)=\overline{X}(T)+\Or(\varepsilon).
\end{equation}
Here
\begin{equation}\label{V.D.13}
\overline{X}(T)=X(0)\cos\sqrt{2}\pi\eta T+\frac{1}{\sqrt{2}\pi\eta}\left.\frac{dX}{dT}\right|_{T=0}\sin\sqrt{2}\pi\eta T
\end{equation}
is the average value of $X=K_1$ as obtained from the averaged system (\ref{V.D.2}), i.e., $\overline{X}=\overline{K}_1$, and the initial data for the Landau decomposition $(X,\xi)$ and its averaged version $(\overline{X},\overline{\xi})$ have been identified. Thus, combined with the compatibility condition (\ref{V.C.5}), (\ref{V.B.11}) and (\ref{V.D.13}) yield the salient consequence that \emph{the original Jacobi field $J^x$ and the oscillation center $X$ in the Landau decomposition $(X,\xi)$ agree in the sense of average:}
\begin{equation}\label{V.D.14}
\overline{J^x}(T)=\overline{X}(T).
\end{equation}
Hence
\numparts
\begin{eqnarray}\label{V.D.15}
J^x(T)&=X(T)+\Or(\varepsilon),\label{V.D.15a}\\
\xi(T)&=\Or(\varepsilon)\label{V.D.15b}
\end{eqnarray}
\endnumparts
uniformly for $0\le T\le\ln(1+\varepsilon)/2\sqrt{2}\pi\eta$ in the sense that
\begin{equation}\label{V.D.16}
\left|J^x(T)-X(T)\right|=\left|\xi(T)\right|\le\left(|J^x(0)|+\frac{1}{\sqrt{2}\pi\eta}\left|\frac{dJ^x}{dT}\right|_{T=0}\right)\varepsilon
\end{equation}
for any $\varepsilon>0$. This quantifies the smallness of the rapid oscillation $\xi$ as compared to the oscillation center $X$.
\subsection{The Ponderomotive Approximation of the Landau Decomposition}
\label{section V.E}

Based on the compatibility condition (\ref{V.C.5}), we will quantify the closeness of the oscillation center $X$ in the Landau decomposition $(X,\xi)$ and its ponderomotive approximation $X_p$, thereby providing a mathematically solid formulation for the ponderomotive oscillation center $X_p$. To this end, we first calculate the ponderomotive oscillation center $X_p$ following \cite{gibbon2005,landau1981,nicholson1983,schmidt1979}. More sophisticated derivations that lead to equivalent results can be found in \cite{bauer1995,startsev1997}. In addition, the result of this calculation leads to an understanding of the breakdown of the ponderomotive approximation when a parametric resonance occurs. This will be expounded in the next section.

While regarding $X$ as {\it constant} over a given optical cycle and taking into account of the zero average of $\xi$ over each optical cycle, we integrate the last equation of (\ref{V.C.4}) to obtain
\begin{equation}\label{V.E.1}
\xi_p(T)=-\frac{\eta^2}{8}\cos4\pi T\cdot X(T).
\end{equation}
Here we have used the subscript `$p$' to distinguish $\xi_p$ obtained this way from the exact rapid small oscillation $\xi$ in the Landau decomposition $(X,\xi)$. Then the evolution equation of $X_p$ is obtained by first replacing $X$ and $\xi$ in the second equation of (\ref{V.C.4}) by $X_p$ and $\xi_p$ respectively and then by averaging the resulting equation. The result is
\begin{equation}\label{V.E.2}
\eqalign{
\frac{d^2X_p}{dT^2}&=-2\pi^2\eta^2X_p-4\pi^2\eta^2\overline{\left(-\frac{\eta^2}{8}\cos 4\pi T\right)\sin^22\pi T}\cdot X_p\cr
&=-2\pi^2\eta^2X_p-\frac{\pi^2\eta^4}{8}X_p\cr
&=-\frac{\pi^2\eta^2(16+\eta^2)}{8}X_p.}
\end{equation}
Consequently,
\begin{equation}\label{V.E.3}
\fl X_p(T)=X(0)\cos\sqrt{2+\frac{\eta^2}{8}}\pi\eta T+\sqrt{\frac{8}{16+\eta^2}}\frac{1}{\pi\eta}\left.\frac{dX}{dT}\right|_{T=0}\sin\sqrt{2+\frac{\eta^2}{8}}\pi\eta T.
\end{equation}
Here the initial data for $X$ and $X_p$ have been identified. Then, in view of (\ref{V.D.13}) and (\ref{V.E.3}), one has
\begin{equation}\label{V.E.4}
|\overline{X}(T)-X_p(T)|\le2\left(\sqrt{2+\frac{\eta^2}{8}}|X(0)|+\frac{1}{\pi\eta}\left|\frac{dX}{dT}\right|_{T=0}\right)\pi\eta T
\end{equation}
by the mean value theorem. This implies that
\begin{equation}\label{V.E.5}
|\overline{X}(T)-X_p(T)|\le\left(\sqrt{2+\frac{\eta^2}{8}}|X(0)|+\frac{1}{\pi\eta}\left|\frac{dX}{dT}\right|_{T=0}\right)\varepsilon
\end{equation}
and thus
\begin{equation}\label{V.E.6}
\overline{X}(T)=X_p(T)+\Or(\varepsilon)
\end{equation}
uniformly for $0\le T\le\varepsilon/2\pi\eta$ for any $\varepsilon>0$. Finally, combining (\ref{V.D.11}) and (\ref{V.E.5}) yields
\begin{equation}\label{V.E.7}
\fl|X(T)-X_p(T)|\le\left[\left(\frac{1}{2}+\sqrt{2+\frac{\eta^2}{8}}\right)|X(0)|+\left(\frac{1}{2\sqrt{2}}+1\right)\frac{1}{\pi\eta}\left|\frac{dX}{dT}\right|_{T=0}\right]\varepsilon,
\end{equation}
showing that
\begin{equation}\label{V.E.8}
X(T)=X_p(T)+\Or(\varepsilon)
\end{equation}
uniformly for $0\le T\le\min\left\{\ln(1+\varepsilon)/2\sqrt{2}\pi\eta,\varepsilon/2\pi\eta\right\}$ for any $\varepsilon>0$. This gives a mathematically rigorous statement about the closeness of the oscillation center $X$ in the Landau decomposition $(X,\xi)$ to its ponderomotive approximation $X_p$, whose evolution is governed by (\ref{V.E.2}) instead, in a sufficiently low impulsive E.M. field ($\eta\ll 1$).
\subsection{Breakdown of Averaged Landau Decomposition and Ponderomotive Approximation due to Parametric Resonance}
\label{section V.F}

We say that the mathematical averaging principle and  the ponderomotive approximation  for the oscillation center $X$ in the Landau decomposition $(X,\xi)$ break down when (\ref{V.D.12}) and (\ref{V.E.8}) respectively become inapplicable as $T\to\infty$. We will exhibit such a breakdown when a parametric resonance occurs in the original Jacobi field $J^x$. This we do by first relating the exponential order of $J^x$ to the ones of $X$ and $\xi$. More explicitly, we show that if any of the members of the Landau decomposition $(X,\xi)$ has a positive exponential order\footnote{If for some $\alpha,M\in{\mathbb R}$, $\left|f(T)\right|\le Me^{\alpha T}$ for all $T\ge 0$, then the function $f$ is said to have the exponential order $\alpha$.}, then so has the other element the same exponential order. Thus it makes sense to talk about the exponential order of the Landau decomposition $(X,\xi)$ instead of the exponential order of any of its elements. Furthermore, the associated Jacobi field $J^x$ possesses the same exponential order as the Landau decomposition $(X,\xi)$. On the other hand, when a parametric resonance occurs, the Jacobi field $J^x$ has exponential order $\mu$ for almost all initial data, where $\mu$ is the positive Floquet exponent associated with the Jacobi equation, i.e., (\ref{V.A.4}). Together with the relation among the exponential orders of the Jacobi field $J^x$ and its associated Landau decomposition $(X,\xi)$, this implies that \emph{both the oscillation center $X$ and the oscillatory component $\xi$, assumed small in the ponderomotive approximation, diverge exponentially in amplitude as $T\to\infty$ with exponent no smaller than the Floquet exponent $\mu$ for almost all initial data in the case when a parametric resonance in $J^x$ takes place.} In fact, both $dX/dT$ and $d\xi/dT$ also diverge exponentially in amplitude as $T\to\infty$ with no smaller exponent under the same circumstances. These divergences suggest an inconsistency with (\ref{V.D.12}) and (\ref{V.E.8}) as $T\to\infty$, where the oscillation center $X$ in the Landau decomposition $(X,\xi)$ is bounded in both its averaged and ponderomotive approximations. In other words, these divergences suggest the breakdown of the mathematical averaging principle and the ponderomotive approximation for $X$. The primary aim of this section is to provide a rigorous mathematical justification of these suspicions.

To begin with, let us relate the exponential orders of the Jacobi field $J^x$ and the members in the associated Laudau decomposition $(X,\xi)$. To this end, note that the evolution equations, i.e., the last two equations of (\ref{V.C.4}), of the Landau decomposition $(X,\xi)$ satisfy the following equivalent integral equations
\begin{equation}\label{V.F.1}
\eqalign{
X(T)&=X(0)\cos\sqrt{2}\pi\eta T+\frac{1}{\sqrt{2}\pi\eta}\left.\frac{dX}{dT}\right|_{T=0}\sin\sqrt{2}\pi\eta T\cr
&\qquad-2\sqrt{2}\pi\eta\int_0^T\,\sin\sqrt{2}\pi\eta(T-S)\sin^22\pi S\cdot\xi(S)\,dS,\label{V.F.1a}\cr
\xi(T)&=\xi(0)+\left.\frac{d\xi}{dT}\right|_{T=0}T+2\pi^2\eta^2\int_0^T\,(T-S)\cos 4\pi S\cdot X(S)\,dS.\label{V.F.1b}}
\end{equation}
First of all, note that these equations imply that \emph{if $X(T)$ has exponential order $\alpha>0$ for $T\ge 0$,  then so have $\xi(T)$ and $J^x(T)$.} To see this, let $M_1\ge 0$ be such that $|X(T)|\le M_1e^{\alpha T}$ for all $T\ge 0$. Then
\begin{equation}\label{V.F.2}
\eqalign{
|\xi(T)|&\le|\xi(0)|+\left|\frac{d\xi}{dT}\right|_{T=0}T+2\pi^2\eta^2M_1\int_0^T\,(T-S)e^{\alpha S}\,dS\cr
&=|\xi(0)|+\left|\frac{d\xi}{dT}\right|_{T=0}T+\frac{2\pi^2\eta^2M_1}{\alpha^2}(e^{\alpha T}-\alpha T-1)\cr
&\le|\xi(0)|+\frac{2\pi^2\eta^2M_1}{\alpha^2}+\left(\left|\frac{d\xi}{dT}\right|_{T=0}+\frac{2\pi^2\eta^2M_1}{\alpha}\right)T\cr
&\qquad+\frac{2\pi^2\eta^2M_1}{\alpha^2}e^{\alpha T}\cr
&\le\left[|\xi(0)|+\frac{1}{\alpha}\left|\frac{d\xi}{dT}\right|_{T=0}+\frac{6\pi^2\eta^2M_1}{\alpha^2}\right]e^{\alpha T}}
\end{equation}
by the second equation of (\ref{V.F.1}), from which it follows that
\begin{equation}\label{V.F.3}
|J^x(T)|\le\left[M_1+|\xi(0)|+\frac{1}{\alpha}\left|\frac{d\xi}{dT}\right|_{T=0}+\frac{6\pi^2\eta^2M_1}{\alpha^2}\right]e^{\alpha T}
\end{equation}
for all $T\ge 0$.

Secondly, if \emph{if $\xi(T)$ has exponential order $\beta>0$ for $T\ge 0$, then so have $X(T)$ and $J^x(T)$.} Explicitly, let $|\xi(T)|\le M_2e^{\beta T}$ for $T\ge 0$, where $M_2\ge 0$. Then
\begin{equation}\label{V.F.4}
\eqalign{
|X(T)|&\le|X(0)|+\frac{1}{\sqrt{2}\pi\eta}\left|\frac{dX}{dT}\right|_{T=0}+2\sqrt{2}\pi\eta\int_0^T\,e^{\beta S}\,dS\cr
&=|X(0)|+\frac{1}{\sqrt{2}\pi\eta}\left|\frac{dX}{dT}\right|_{T=0}+\frac{2\sqrt{2}\pi\eta}{\beta}(e^{\beta T}-1)\cr
&\le|X(0)|+\frac{1}{\sqrt{2}\pi\eta}\left|\frac{dX}{dT}\right|_{T=0}+\frac{2\sqrt{2}\pi\eta}{\beta}(e^{\beta T}+1)\cr
&\le\left[|X(0)|+\frac{1}{\sqrt{2}\pi\eta}\left|\frac{dX}{dT}\right|_{T=0}+\frac{4\sqrt{2}\pi\eta}{\beta}\right]e^{\beta T}}
\end{equation}
and
\begin{equation}\label{V.F.5}
\left|J^x(T)\right|\le\left[M_2+|X(0)|+\frac{1}{\sqrt{2}\pi\eta}\left|\frac{dX}{dT}\right|_{T=0}+\frac{4\sqrt{2}\pi\eta}{\beta}\right]e^{\beta T}
\end{equation}
for all $T\ge 0$.

We are now ready to employ (\ref{V.F.3}) and (\ref{V.F.5}) to justify the statement that \emph{when a parametric resonance occurs in the original Jacobi field $J^x$,} i.e., when the characteristic function $\varphi(\eta)$ of (\ref{V.A.4}) has an absolute value $|\varphi(\eta)|>1$, \emph{both $X$ and $\xi$ diverge exponentially in amplitude as $T\to\infty$ with exponent no smaller than $\mu$ for almost all initial data for $J^x$, where $\mu$ is the positive Floquent exponent associated with the Jacobi equation (\ref{V.A.4}).} As we have pointed out earlier, \emph{this suggests the breakdown of the mathematical averaging principle and the ponderomotive approximation for the oscillation center $X$ because there both the averaged and the ponderomotive approximations of $X$ remain bounded for all time.} 

To demonstrate the exponential divergences of $X$ and $\xi$ in the case of parametric resonance, recall from standard Floquet theory that there is a fundamental set of Jacobi fields that take the Bloch-Floquet forms $J_1(T)=e^{\mu T}K_1(T)$ and $J_2(T)=e^{-\mu T}K_2(T)$, where $K_1$ and $K_2$ are periodic functions with the optical cycle being their common period \cite{chicone2006}. Let $J^x=c_1J_1+c_2J_2$, where $c_1$ and $c_2$ are integration constants uniquely determined by the initial data for $J^x$ and vice verse. Thus $J^x$ diverges exponentially at the rate of the Floquet exponent $\mu$ for almost all initial data. Let $0\le T_1<1$ be such that $K_1(T_1)\ne 0$. Then for any integer $N>0$,
\begin{equation}\label{V.F.6}
J^x(T_1+N)=c_1K_1(T_1)e^{\mu(T_1+N)}+c_2K_2(T_1)e^{-\mu(T_1+N)}.
\end{equation}
Consequently,
\begin{equation}\label{V.F.7}
|c_1K_1(T_1)|e^{\mu(T_1+N)}\le|J^x(T_1+N)|+|c_2K_2(T_1)|e^{-\mu(T_1+N)}.
\end{equation}
Hence if $X(T)$ has exponential order $\alpha>0$ for $T\ge 0$, then it follows from (\ref{V.F.3}) and (\ref{V.F.7}) that
\begin{equation*}
|c_1K_1(T_1)|e^{\mu(T_1+N)}\le M_3e^{\alpha(T_1+N)}+|c_2K_2(T_1)|e^{-\mu(T_1+N)}
\end{equation*}
or equivalently
\begin{equation*}
|c_1K_1(T_1)|\le M_3e^{(\alpha-\mu)(T_1+N)}+|c_2K_2(T_1)|e^{-2\mu(T_1+N)}
\end{equation*}
for some $M_3\ge 0$. If $\alpha<\mu$, this clearly leads to a contradiction as we pass to the limit $N\to\infty$ unless $c_1=0$. Thus the oscillation center $X$ must diverge exponentially faster than its associated Jacobi field $J^x$. Indeed, the oscillatory component $\xi$ must also diverge exponentially at a rate larger than than of $J^x$. To see this, suppose $\xi$ has exponential order $\beta>0$ for $T\ge 0$, then it follows from (\ref{V.F.5}) and (\ref{V.F.7}) that
\begin{equation*}
|c_1K_1(T_1)|e^{\mu(T_1+N)}\le M_4e^{\beta(T_1+N)}+|c_2K_2(T_1)|e^{-\mu(T_1+N)}
\end{equation*}
or equivalently
\begin{equation*}
|c_1K_1(T_1)|\le M_4e^{(\beta-\mu)(T_1+N)}+|c_2K_2(T_1)|e^{-2\mu(T_1+N)}
\end{equation*}
for some $M_4\ge 0$. If $\beta<\mu$, we also reach a contradiction by sending $N\to\infty$ unless $c_1=0$.

To conclude this section, we note that the point of departure to validate the exponential divergences of $dX/dT$ and $d\xi/dT$ can be taken to be the equations
\begin{equation}\label{V.F.8}
\eqalign{
\frac{dX}{dT}&=-\sqrt{2}\pi\eta X(0)\sin\sqrt{2}\pi\eta T+\left.\frac{dX}{dT}\right|_{T=0}\cos\sqrt{2}\pi\eta T\cr
&\quad-4\pi^2\eta^2\int_0^T\,\cos\sqrt{2}\pi\eta(T-S)\sin^22\pi S\cdot\xi(S)\,dS,\label{V.F.8a}\\
\frac{d\xi}{dT}&=\left.\frac{d\xi}{dT}\right|_{T=0}+2\pi^2\eta^2\int_0^T\,\cos 4\pi S\cdot X(S)\,dS\label{V.F.8b}}
\end{equation}
obtained by differentiating (\ref{V.F.1}). Then precisely the same line of argument used to show the exponential divergences of $X$ and $\xi$ from (\ref{V.F.1}) leads one from (\ref{V.F.8a}) to the conclusion that both $dX/dT$ and $d\xi/dT$ diverge exponentially faster than $dJ^x/dT$ for almost all initial data for $J^x$ in the case of a parametric resonance.

\begin{figure}
\centering
\subfigure[$\eta=0.2$, $\varphi(\eta)\approx 0.902673$]{
\includegraphics[scale =0.36]{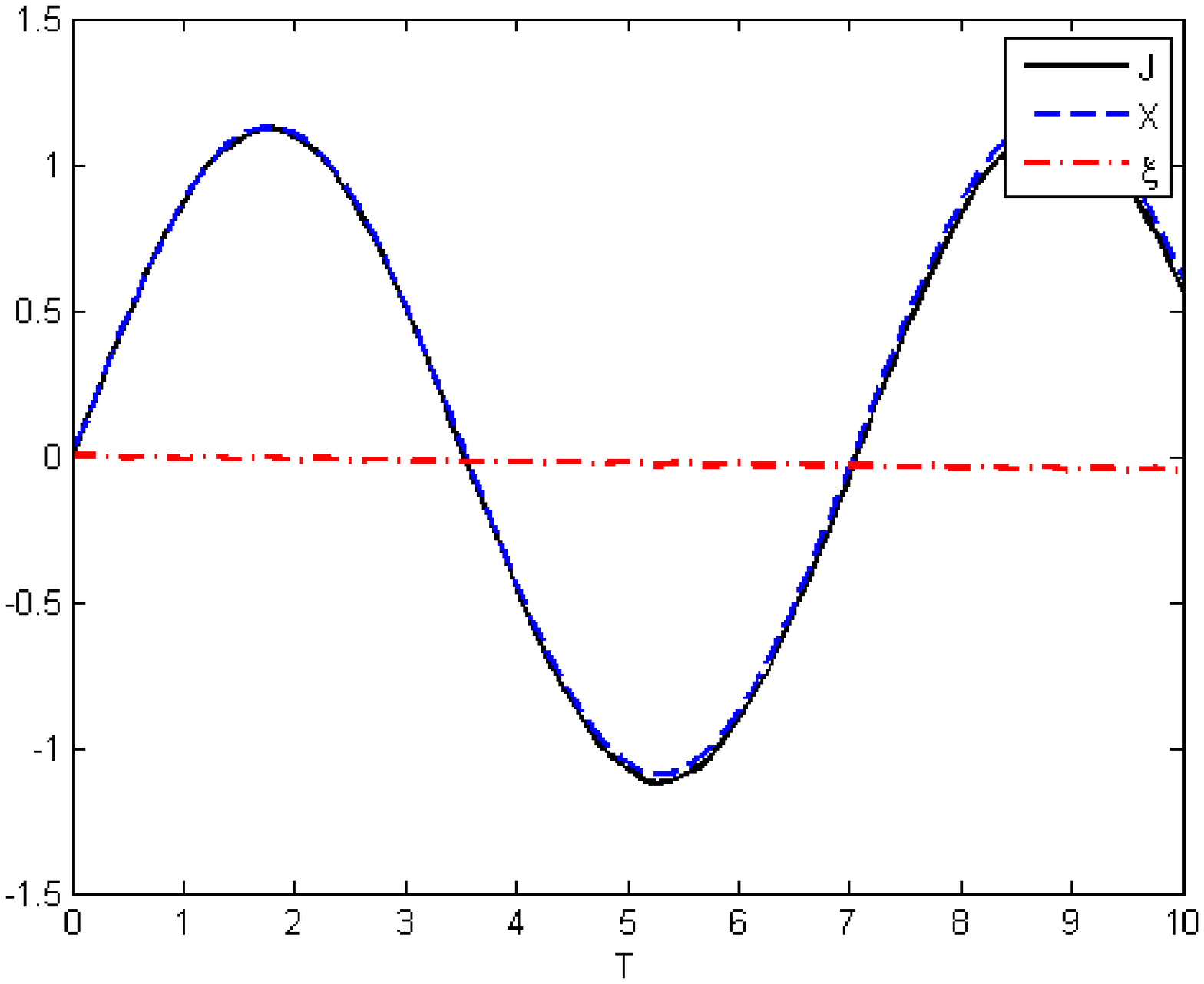}
\label{fig:subfig1}
}
\subfigure[$\eta=0.2$, $\varphi(\eta)\approx 0.902673$]{
\includegraphics[scale =0.36]{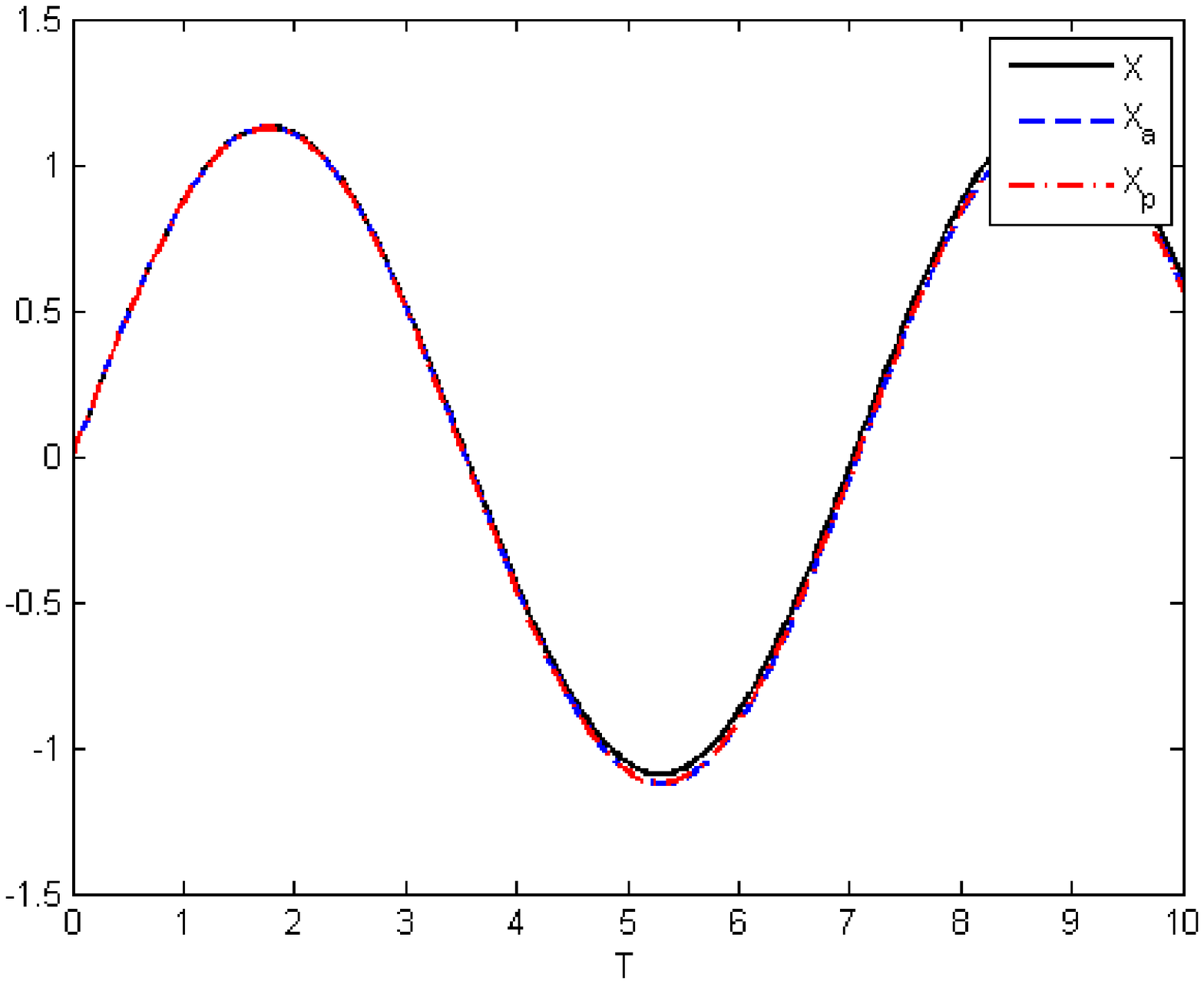}
\label{fig:subfig2}
}
\subfigure[$\eta=0.5$, $\varphi(\eta)\approx  0.435131$]{
\includegraphics[scale =0.36]{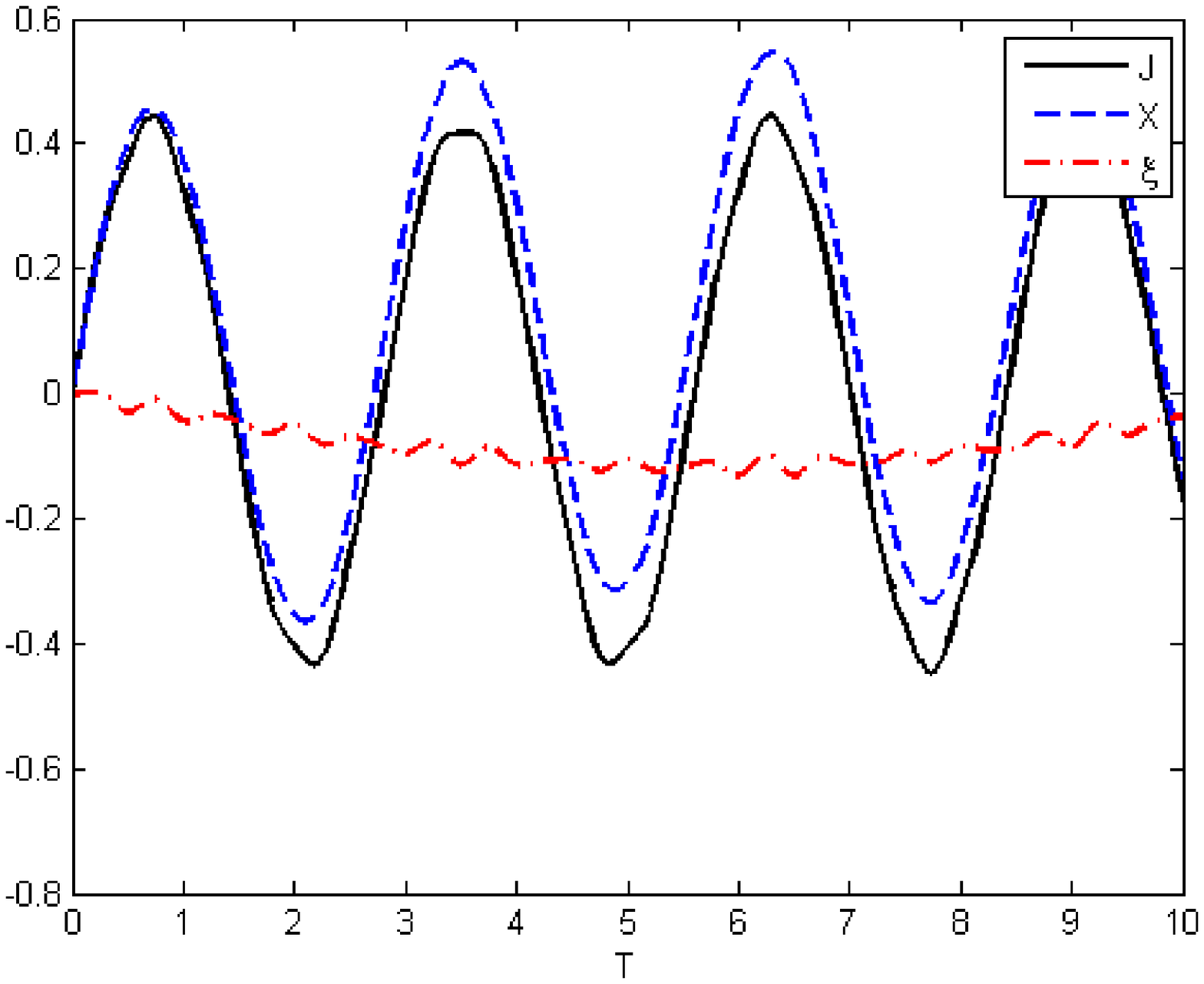}
\label{fig:subfig3}
}
\subfigure[$\eta=0.5$, $\varphi(\eta)\approx  0.435131$]{
\includegraphics[scale =0.36]{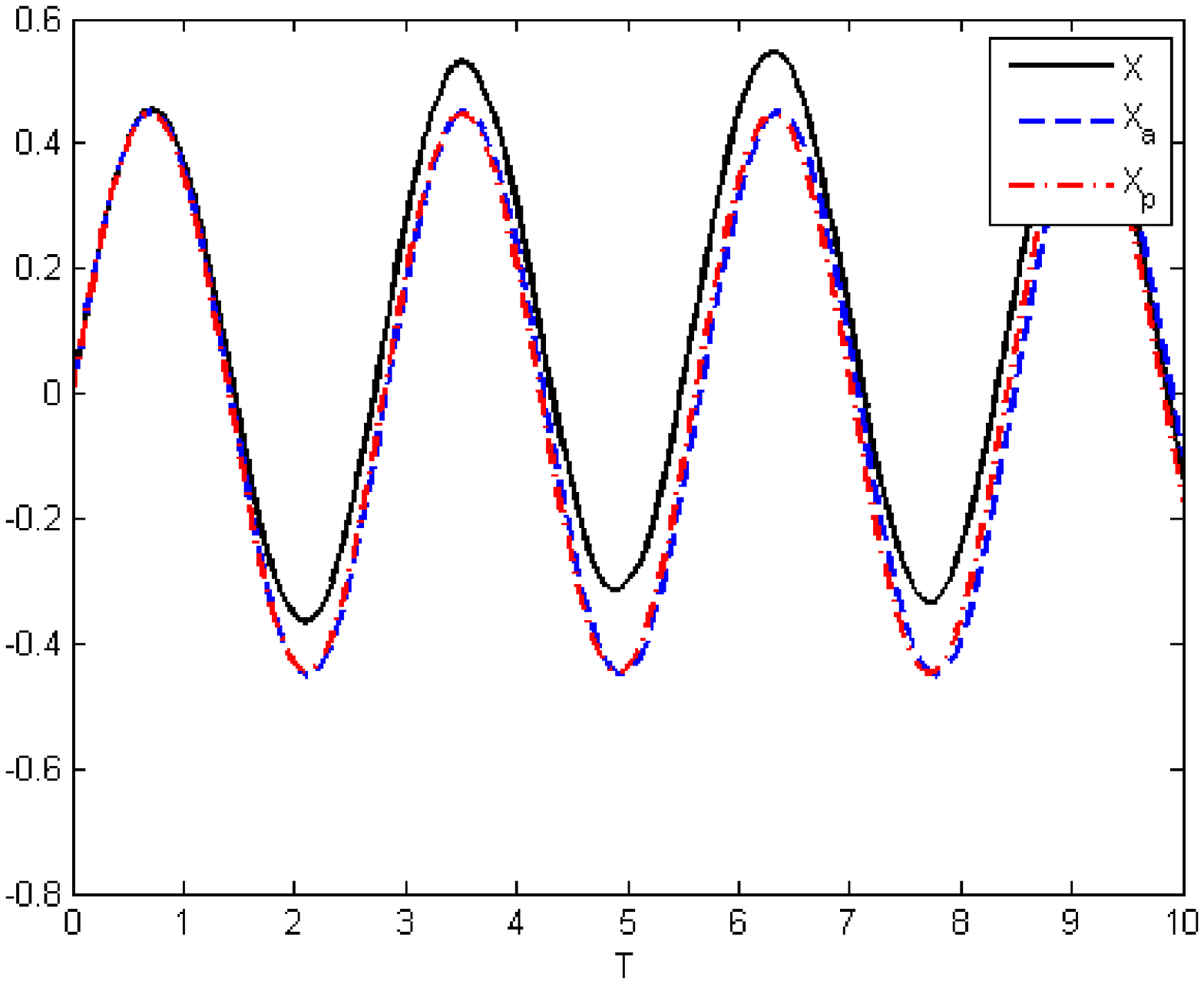}
\label{fig:subfig4}
}
\subfigure[$\eta=1.2$, $\varphi(\eta)\approx  -1.084503$]{
\includegraphics[scale =0.36]{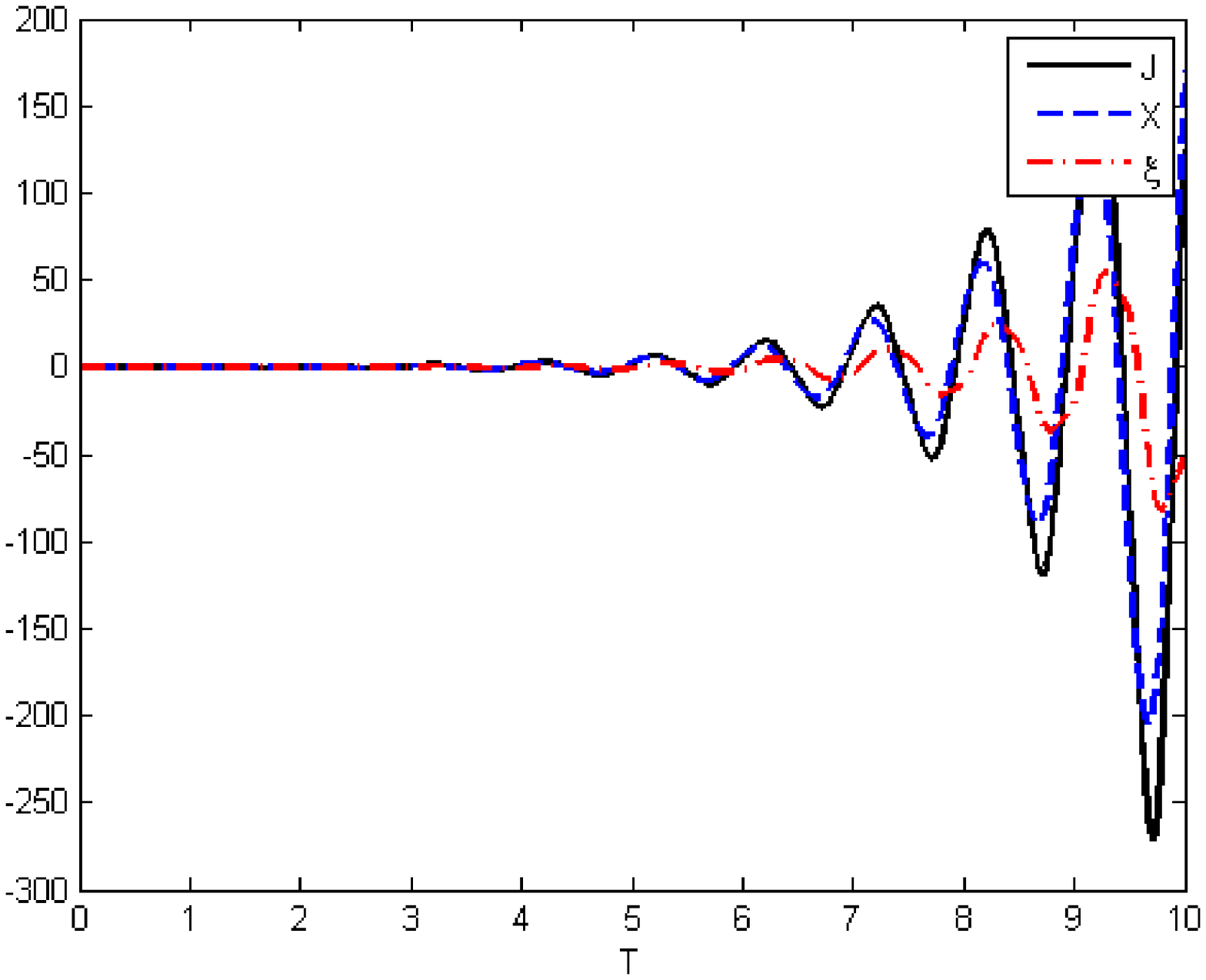}
\label{fig:subfig5}
}
\subfigure[$\eta=1.2$, $\varphi(\eta)\approx  -1.084503$]{
\includegraphics[scale =0.36]{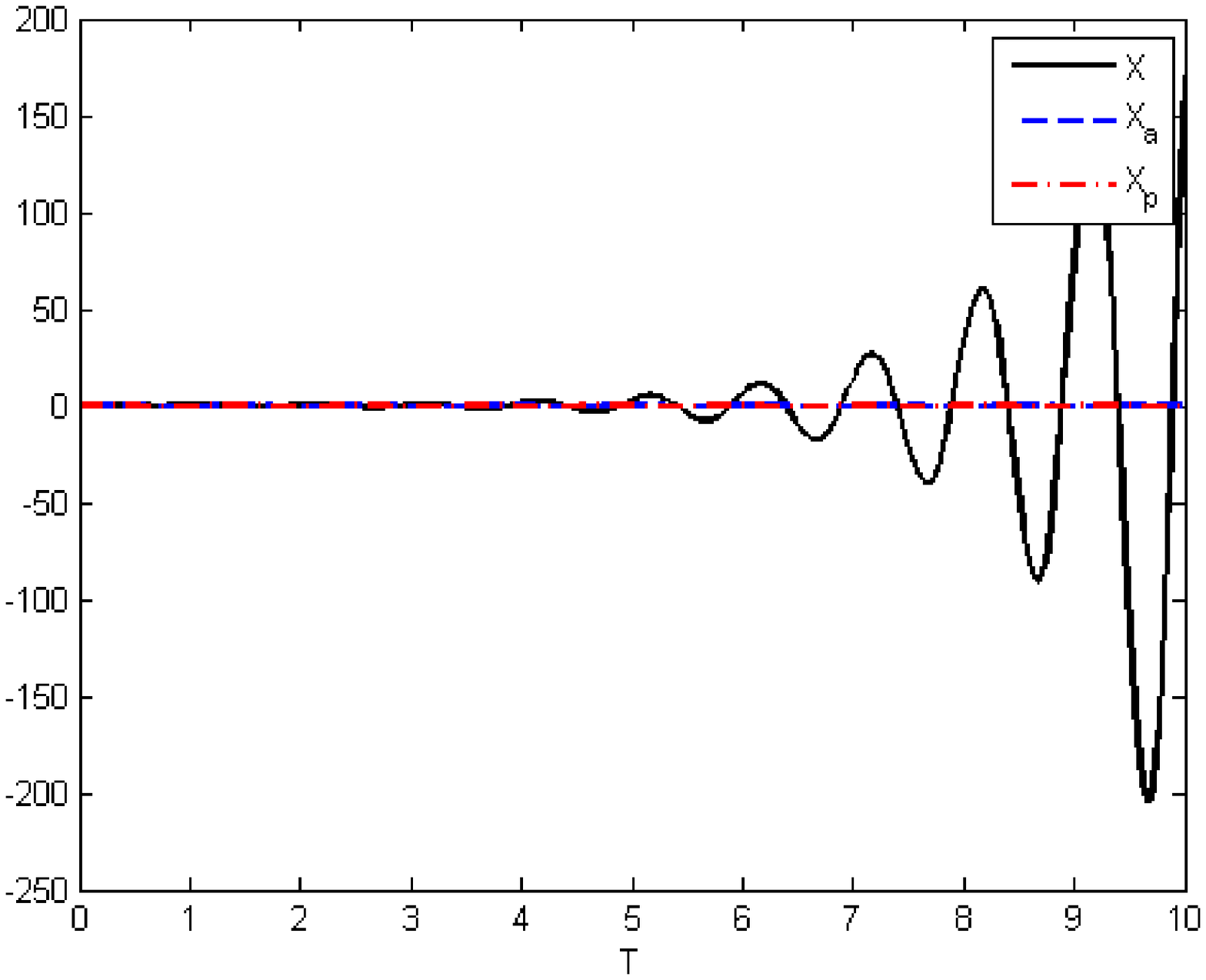}
\label{fig:subfig6}
}
\label{fig:figure1}
\caption[]{The evolution of the Jacobi field $J^x$ and its associated Landau decomposition $(X,\xi)$ as well as the oscillation center $X$ and its averaged and ponderomotive approximations $\overline{X}=X_a$ and $X_p$ respectively for $10$ optical cycles for different values
  of $\eta$. The initial data are $J^x=1$ and $dJ^x/dT=0$. When $\eta\ll 1$,  $J^x\approx X\approx X_a\approx X_p$ for many cycles. When $\eta=1.2$, the Jacobi field $J^x$ experiences a parametric resonance. In this case, both $X$ and $\xi$ diverge exponentially faster than $J^x$. In consequence, neither $X_a$ nor $X_p$ serves as a reasonble approximation to $X$.}
\end{figure}
\section{Conclusions}
\label{section VI}
 
In this paper,

\begin{enumerate}
\item we have reformulated and analyzed the dynamical properties of the orbits of charged particles interacting with a generic E.M. field with planar symmetry, as governed by the Lorentz equations of motion, within a differential geometric framework whose geodesics suffer a mutual deviation in accordance with a curvature induced by the field intensity;

\item we have demonstrated the integrability of the particle motion in a plane-wave field as a consequence of the vanishing of the curvature;

\item  we have indicated the methodology of examining the local stability of the orbit of a particle through the Jacobi field within such a geometrical formulation; and

\item we have also showed the relevance of the geometrical formulation in discussing the domain of applicability of the ponderomotive oscillation center of a particle executing oscillatory motion in the E.M. field.
\end{enumerate}

In addition, by considering the motion of a charged particle in a linearly polarized standing wave field, we have shown that

\begin{enumerate}[(a)]
\item different numerical subintervals of the impulse factor $\eta$ give rise to stable or unstable orbits as a consequence of the absence or the presence of parametric resonance of the Jacobi field;

\item in a sufficiently low impulsive E.M. field ($\eta\ll 1$), a mathematically rigorous averaging principle can be applied to provide a precise meaning to the ponderomotive oscillation center of an orbit;

\item in the occurrence of parametric resonance, the applicability of the ponderomotive approximation breaks down.
\end{enumerate}

These are very interesting and promising results towards an in-depth differential geometric analysis for the dynamics of radiation-particle interaction.

\ack{The author wishes to acknowledge fruitful and interesting discussions with Professor Ulrich H. Gerlach.

\section*{References}
\bibliographystyle{unsrt}
\bibliography{radiationparticle}

\end{document}